\documentclass[twocolumn]{aastex631}

\usepackage{tabularx}
\usepackage{newtxtext,newtxmath}
\usepackage[T1]{fontenc}
\usepackage{nicefrac}
\usepackage{amssymb, amsmath, bm}
\usepackage{mathrsfs}
\usepackage{url}
\usepackage{cases}
\usepackage{epsfig}
\usepackage{color}
\usepackage{graphicx}
\usepackage{grffile}
\usepackage{float}
\usepackage{soul}
\usepackage{enumerate}
\usepackage{multirow}
\usepackage{threeparttable}
\usepackage{xspace}
\usepackage{etoolbox}

\newcommand{\blue}[1]{\textcolor{blue}{{#1}}}

\usepackage{xcolor}
\definecolor{green}{RGB}{50,205,50}

\begin{document}

\title{A Practical Framework for Estimating the Repetition Likelihood of Fast Radio Bursts from Spectral Morphology}

\author[0000-0001-5950-7170]{Wan-Peng Sun}
\affiliation{Liaoning Key Laboratory of Cosmology and Astrophysics, College of Sciences, Northeastern University, Shenyang 110819, China}

\author[0000-0002-8744-3546]{Yong-Kun Zhang}
\affiliation{National Astronomical Observatories, Chinese Academy of Sciences, Beijing 100101, China}

\author[0009-0004-5607-9181]{Ji-Guo Zhang}
\affiliation{Liaoning Key Laboratory of Cosmology and Astrophysics, College of Sciences, Northeastern University, Shenyang 110819, China}

\author[0000-0002-2552-7277]{Xiaohui Liu}
\affiliation{National Astronomical Observatories, Chinese Academy of Sciences, Beijing 100101, China}
\affiliation{School of Astronomy and Space Science, University of Chinese Academy of Sciences, Beijing 100049, China}

\author[0000-0003-1962-2013]{Yichao Li}
\affiliation{Liaoning Key Laboratory of Cosmology and Astrophysics, College of Sciences, Northeastern University, Shenyang 110819, China}

\author[0000-0002-5936-8921]{Fu-Wen Zhang}
\affiliation{College of Science, Guilin University of Technology, Guilin 541004, China}

\author[0009-0004-8194-7446]{Wan-Ting Hou}
\affiliation{College of Mathematics and Statistics, Liaoning University, Shenyang 110036, China}

\author[0000-0002-3512-2804]{Jing-Fei Zhang}
\affiliation{Liaoning Key Laboratory of Cosmology and Astrophysics, College of Sciences, Northeastern University, Shenyang 110819, China}

\author[0000-0002-6029-1933]{Xin Zhang}
\affiliation{Liaoning Key Laboratory of Cosmology and Astrophysics, College of Sciences, Northeastern University, Shenyang 110819, China}
\affiliation{MOE Key Laboratory of Data Analytics and Optimization for Smart Industry, Northeastern University, Shenyang 110819, China}
\affiliation{National Frontiers Science Center for Industrial Intelligence and Systems Optimization, Northeastern University, Shenyang 110819, China}

\correspondingauthor{Yichao Li}
\email{liyichao@mail.neu.edu.cn}
\correspondingauthor{Fu-Wen Zhang}
\email{fwzhang@pmo.ac.cn}
\correspondingauthor{Xin Zhang}
\email{zhangxin@mail.neu.edu.cn}

\begin{abstract}
The repeating behavior of fast radio bursts (FRBs) is regarded as a key clue to understanding their physical origin, yet reliably distinguishing repeaters from apparent non-repeaters with current observations remains challenging. Here we propose a physically interpretable and practically quantifiable classification framework based on spectral morphology. Using dimensionality reduction, clustering, and feature-importance analysis, we identify the spectral running $r$ and spectral index $\gamma$ as the most critical parameters for distinguishing repeaters from apparent non-repeaters in the CHIME/FRB sample. In the $\gamma$-$r$ space, repeaters preferentially occupy regions with steeper, narrower-band spectra, whereas non-repeaters cluster in flatter, broader-band regions, resulting in a clear density separation. We further construct a probability map in the $\gamma$-$r$ space based on Gaussian Mixture Model posterior analysis, revealing a clear gradient of repetition likelihood from $\sim$67\% in the high-repetition region to $\sim$7\% in the low-repetition region. This model also identifies several apparent non-repeaters with high inferred repetition probability, highlighting them as priority targets for future monitoring. This framework provides a simple and generalizable tool for assessing repeatability in the CHIME/FRB sample and highlights the diagnostic power of spectral morphology in unveiling FRB origins.
\end{abstract}


\section{Introduction} \label{sec:intro}
Fast radio bursts (FRBs) are millisecond-duration radio transients of extraordinary luminosity originating from cosmological distances \citep{2007Sci...318..777L,2013Sci...341...53T,2014ApJ...790..101S,2017ApJ...834L...7T,2017ApJ...834L...8M,2022A&ARv..30....2P,2023RvMP...95c5005Z,2024ARNPS..74...89Z}.  
Current observations indicate that FRBs originate at least from compact objects and the burst detected from the Galactic magnetar SGR 1935+2154 has confirmed magnetars as a plausible origin for at least a subset of FRBs \citep{2020Natur.587...59B,2020Natur.587...54C}. Although their physical origins remain elusive, FRBs have emerged as powerful tools for probing the cosmic content \citep{2020Natur.581..391M,2020ApJ...903...83Z,2023SCPMA..6620412Z,2025SCPMA..6880406Z,2025arXiv250706841Z}. FRBs are commonly classified as repeaters, which produce multiple events, or apparent non-repeaters with only a single detected event \citep{2016Natur.531..202S,2021ApJS..257...59C,2023ApJ...947...83C}. However, whether all FRBs will eventually repeat or share a common progenitor remains a central and unresolved question in FRB research \citep{2023RvMP...95c5005Z}.

Previous studies have revealed statistical differences between repeating and apparently non-repeating FRBs across multiple observational properties. Repeaters typically exhibit longer pulse durations, narrower bandwidths \citep{Ziggy2021,2021MNRAS.500.2525K,2023ApJ...947...83C}, and a pronounced downward frequency drift (the `sad trombone'), whereas non-repeaters tend to manifest as shorter, broadband bursts \citep{2021ApJS..257...59C,Ziggy2021,2021MNRAS.500.3275C,2022Univ....8..355Z}. These contrasts were once taken as evidence for fundamentally distinct physical origins of the two populations. However, as observational samples have grown, these distinctions have blurred, as many repeaters now exhibit broader bandwidths and shorter durations \citep{2023ApJ...947...83C,2025ApJ...988...41Z}, which were previously considered typical of non-repeaters. This trend has led to an increasing overlap between the two populations in feature space. Furthermore, after correcting for exposure time and sensitivity, \citet{2023ApJ...947...83C} reported no clear bimodality in burst-rate upper limits between repeaters and non-repeaters, leaving open the possibility that all FRBs may eventually repeat. Further evidence points to intrinsic links between the two populations, particularly in their energy output, suggesting that repeaters and non-repeaters may arise from statistically similar underlying populations. For instance, \citet{2024NatAs...8..337K} found that the high-energy bursts of the hyperactive repeater FRB 20201124A closely align with the energy distribution of the overall non-repeater population, implying that apparently non-repeating FRBs could simply be rare and exceptionally bright events from repeating sources. Similarly, \citet{2024arXiv241017024O} reported a consistent trend at the high-energy end for both populations. In other words, despite differences in properties such as pulse width and bandwidth, their energy-release behaviors notably overlap \citep{2021ApJ...920L..23Z,2025RAA....25h5009H}. These findings challenge the notion that FRBs comprise two fundamentally distinct classes and highlight the difficulties in classifying them based on existing observational criteria. Nevertheless, refining their separation in parameter space and identifying potential repeater candidates remain crucial for advancing FRB studies.

In recent years, numerous studies have attempted to distinguish repeaters from apparent non-repeaters using 
supervised machine learning approaches \citep{2023MNRAS.518.1629L,2024MNRAS.533.3283S,Kharel:2025oza} and 
unsupervised \citep{2022MNRAS.509.1227C,2023MNRAS.522.4342Y,2023MNRAS.519.1823Z,2024ApJ...977..273G,Sun2025,2025ApJ...982...16Q,Liu2025,2025PASP..137l4102M,2026JHEAp..4900449J}. 
Generally, the supervised approaches face the primary challenge of avoiding erroneous training and selecting representative, well-labeled datasets for robust model development. Given current observational limitations, inappropriate training sets or biases can lead to misleading classifications, necessitating caution in their application. 

The unsupervised methods are particularly valuable for leveraging multidimensional observational parameters for dimensionality reduction and clustering. However, feature selection is critical for unsupervised clustering, and current studies exhibit substantial variability in the choice of input feature parameters. Because models differ in their sensitivity to input parameters, current studies have produced inconsistent clustering results and identified varying sets of key distinguishing features, making it difficult to evaluate the reliability of any single analysis.
In principle, if there truly exist key physical parameters that 
optimally distinguish repeaters from non-repeaters, independent methodologies should converge upon comparable separation criteria. The fact that current studies instead highlight different key physical features reflects the limited robustness of these approaches. 

Overall, existing machine learning methods often rely on high-dimensional inputs and complex models, limiting their physical interpretability and generalizability. Moreover, their outputs are typically confined to binary classifications without accompanying probabilistic assessments, thereby constraining their practical utility for prioritizing and guiding follow-up observations.

To address these challenges, we propose a physically interpretable and statistically robust framework that estimates the probability of an FRB exhibiting repeater-like properties, thereby providing a probabilistic metric to inform follow-up strategies. In this work, we perform a systematic analysis of multidimensional observational parameters using an extended CHIME/FRB sample (combining the first CHIME/FRB catalog \citep{2021ApJS..257...59C} with newly reported repeaters from \citet{2023ApJ...947...83C}). The structure of our paper is as follows: In Section \ref{sec:data}, we introduce the extended CHIME/FRB sample and the methods used in this work. Section \ref{sec:resu} presents the clustering results and the discussions. In Section \ref{sec:concl}, we summarize our work.

\section{DATA AND ALGORITHM PARAMETERS} \label{sec:data}

\subsection{Data Sample Selection} \label{subsec:data sel}
This study draws on FRB observations from two samples:
\begin{enumerate}
\item The first CHIME/FRB catalog (hereafter Catalog 1) comprises FRBs observed between 2018 July 25 and 2019 July 1 \citep{2021ApJS..257...59C}.
\item The CHIME/FRB Collaboration (2023) catalog (hereafter Catalog 2023) includes repeating FRBs detected from 2019 September 30 to 2021 May 1 \citep{2023ApJ...947...83C}.
\end{enumerate}

All data were recorded within a 400--800 MHz frequency range. Each FRB may consist of one or more pulses that appear as isolated peaks in the dynamic spectrum, referred to as sub-bursts \citep{2024MNRAS.52710425S,2024MNRAS.529L.152B}. Since different sub-bursts may exhibit distinct spectral characteristics and parameter values, each sub-burst is treated as an independent event in our analysis.

We merge the FRB samples from Catalog 1 with the confirmed repeating FRBs from Catalog 2023. We exclude six non-repeating FRBs that lack flux measurements, as well as one burst from FRB 20210224A (Sub\_num = 0) due to its large uncertainty in $r$, potentially compromising the reliability of subsequent analysis. The final dataset consists of 706 FRBs, including 494 non-repeaters and 212 repeaters from 43 distinct sources.

\subsection{FRB Feature Parameters} \label{ssec:data desc}
In our previous work \citep{Sun2025}, we selected seven key observational parameters to characterize the intrinsic emission properties and morphological characteristics of FRBs. These include the flux ($S_{\nu}$), fluence ($F_{\nu}$), sub-burst pulse width ($\Delta t_\mathrm{WS}$), scattering time ($\Delta t_\mathrm{ST}$), spectral index ($\gamma$), spectral running ($r$), and the lowest frequency ($\nu_\mathrm{Low}$). 
Perticularly, the spectral index $\gamma$ and spectral running $r$ well characterize the 
spectral morphology of a single burst via \citep{2021ApJS..257...59C}:
\begin{equation}
  S(\nu) = S_0 \left( \frac{\nu}{\nu_0} \right)^{\gamma + r \ln \left( \frac{\nu}{\nu_0} \right)},
  \label{equ:equ1}
\end{equation}
where $S(\nu)$ is the flux at frequency $\nu$, $S_0$ denotes the overall amplitude and $\nu_0 = 400.2\,{\rm MHz}$
is a selected reference frequency. 
The spectral index $\gamma$ controls the overall slope of the spectrum, while $r$ introduces curvature by modulating the exponent logarithmically.
These features are selected to capture the essential differences between repeating and non-repeating FRBs in terms of their radiation mechanisms and propagation environments.

In the present study, we extend the feature set by incorporating two additional parameters related to spectral structure: the highest frequency ($\nu_\mathrm{High}$) and the peak frequency ($\nu_\mathrm{Peak}$). While $\nu_\mathrm{High}$ was previously excluded to avoid potential biases introduced by truncation at CHIME’s upper frequency limit, we now reconsider such parameter in light of accumulated data, aiming to more comprehensively characterize the spectral extent of FRBs within the 400--800 MHz band. Notably, the truncation itself may encode meaningful differences between the repeaters and non-repeaters.
The parameter $\nu_\mathrm{Peak}$ is defined as the frequency at which each burst reaches its peak intensity within the observed band. Given the diversity in FRB spectral morphology, this parameter is expected to enhance the discriminative power of subsequent classification analyses.

Thus, we selecte a total of nine feature parameters in this study: $S_{\nu}$, $F_{\nu}$, $\Delta t_\mathrm{WS}$, $\Delta t_\mathrm{ST}$, $\gamma$, $r$, $\nu_\mathrm{Low}$, $\nu_\mathrm{High}$, and $\nu_\mathrm{Peak}$.

\subsection{t-SNE Algorithm} \label{ssec:algor}
In this study, we apply the t-SNE \citep[t-distributed Stochastic Neighbor Embedding,][]{JMLR:v9:vandermaaten08a,JMLR:v15:vandermaaten14a} algorithm to perform nonlinear dimensionality reduction on the multi-dimensional observational parameters of FRBs, enabling visualization of their similarity structure in a two-dimensional space. The t-SNE algorithm converts high-dimensional distances into conditional probability distributions to preserve local neighborhood relationships, making it well-suited for revealing potential distribution patterns and cluster structures in high-dimensional data.

The quality of t-SNE embeddings depends primarily on three key hyperparameters: perplexity, early exaggeration, and learning rate. Among these, perplexity controls the trade-off between preserving local and global structure and is generally regarded as the most significant parameter. Early exaggeration enhances initial similarities to facilitate well-separated clusters, while the learning rate controls the step size in gradient descent, affecting the scale and separation of points in the low-dimensional embedding. A detailed description of these parameters can be found in the Scikit-learn documentation\footnote{\url{https://scikit-learn.org/stable/modules/generated/sklearn.manifold.TSNE.html}}.

\begin{table}
\scriptsize
\begin{center}
\begin{threeparttable}
\caption{Sensitivity of t-SNE embeddings to the choice of perplexity, evaluated using trustworthiness, continuity, and Spearman correlation.}
\label{tab:tab1}
\tabletypesize{\scriptsize}
\setlength{\tabcolsep}{3pt}
\begin{tabular}{cccc}
\hline
Perplexity    & Trustworthiness  & Continuity  & Spearman correlation\\
\hline
40  & 0.998 & 0.743 & 0.722 \\
50  & 0.998 & 0.748 & 0.736 \\
60  & 0.998 & 0.752 & 0.712 \\
70  & 0.997 & 0.744 & 0.732 \\
80  & 0.997 & 0.745 & 0.711 \\
\hline
\end{tabular}
\begin{tablenotes}
\item \textbf{Note.} Trustworthiness and continuity measure the preservation of local and global neighborhood structures in the low-dimensional embedding, while the Spearman correlation quantifies the consistency of pairwise distances between the original feature space and the embedding. The small variations across different perplexity values indicate the robustness of the t-SNE results, supporting the choice of perplexity = 60 for subsequent analyses.
\end{tablenotes}
\end{threeparttable}
\end{center}
\end{table}

In this work, the early exaggeration
and learning rate are fixed to their commonly used values of 12 and 260, respectively, which are consistent with empirically recommended ranges and were found to yield stable embeddings in preliminary tests. To assess the robustness of the t-SNE embeddings, we vary the perplexity
between 40 and 80 and evaluate stability using three quantitative metrics:
trustworthiness, continuity, and Spearman correlation. 
The corresponding results are listed in Table~\ref{tab:tab1}. 
Trustworthiness measures the fraction of neighbors in the embedding
that are also neighbors in the original high-dimensional space, 
while continuity quantifies the fraction of true neighbors 
that are preserved in the embedding \citep{trustworthiness}. 
High values of both indicate good local structure preservation. 
The Spearman correlation further tests the monotonic relation between
pairwise distances in the original and embedded spaces \citep{kokoska2000},
providing a complementary global measure of fidelity. 
Across this range of perplexity values, all three metrics remain stable,
indicating that the embeddings are robust. We therefore adopt a perplexity of 60 for the final results used in subsequent classification. It is worth noting that the t-SNE embedding coordinates themselves have no direct physical meaning; however, their relative positions encode the similarity relationships among FRBs in the original feature space.

\section{RESULTS AND DISCUSSION} \label{sec:resu}
\subsection{t-SNE Results} \label{subsec:class}

\begin{figure}
\centering
\includegraphics[angle=0,scale=0.58]{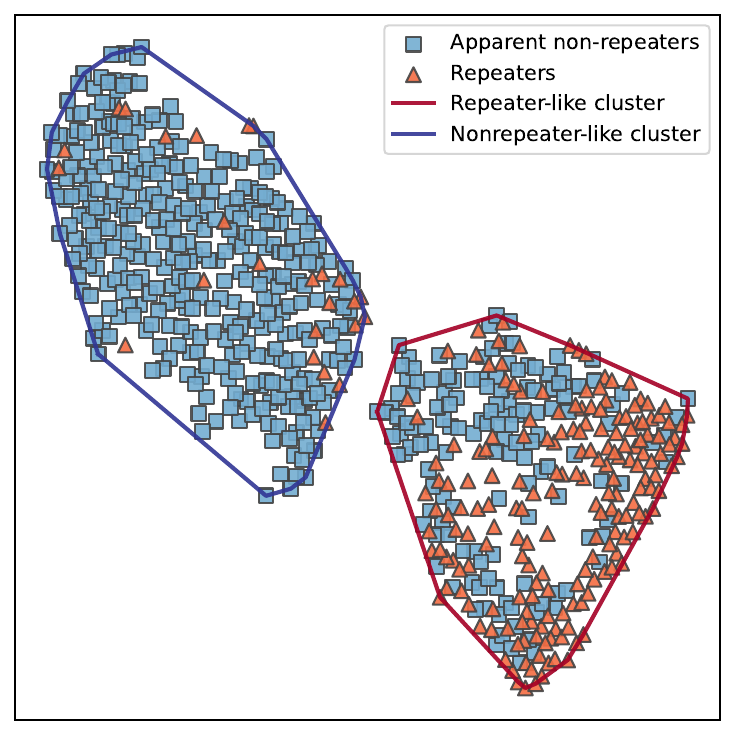}
\caption{
The embedding space of the t-SNE dimension reduction results and HDBSCAN clustering of FRBs in the CHIME/FRB Catalog 1 and Catalog 2023. Blue squares represent apparent non-repeaters, while red triangles denote repeaters. The blue and red contours indicate the clusters identified by HDBSCAN, corresponding to the Nonrepeater-like cluster and Repeater-like cluster, respectively.
\label{fig:figure1}}
\end{figure}

Figure \ref{fig:figure1} presents the t-SNE embedding result for FRBs in both Catalog 1 and Catalog 2023. Red triangles denote repeaters, while blue squares represent apparent non-repeaters. Overall, the FRBs exhibit a distinct bimodal clustering in the t-SNE embedding space, with most repeaters clustered in the lower-right region and non-repeaters predominantly in the upper-left. Since t-SNE preserves local neighborhood relations, the observed separation likely reflects two dominant combinations of features, with certain regions of the parameter space (e.g., the lower-right cluster) preferentially associated with repeaters. While the axes are not physically interpretable, this clustering pattern highlights a statistically significant association between FRB repetition behavior and their underlying properties.

To further characterize the structure revealed by the t-SNE embedding, we apply the HDBSCAN  \citep[Hierarchical Density-Based Spatial Clustering of Applications with Noise,][]{McInnes2017} algorithm for unsupervised, density-based clustering to extract statistically robust subgroups of FRBs.
HDBSCAN is a density-based clustering method that does not require predefining the number of clusters and is well-suited for identifying statistically significant groupings in complex, non-spherical structures, particularly within nonlinear embeddings such as those produced by t-SNE. As shown in Figure \ref{fig:figure1}, the HDBSCAN identifies two primary clusters, delineated by blue and red contour lines, respectively. To highlight their contrasting tendencies in repeating activity, we designate the cluster that includes the majority of repeating FRBs as the Repeater-like cluster, emphasizing its strong statistical association with repeating FRB behavior. The cluster, primarily composed of non-repeating FRBs, is referred to as the Nonrepeater-like cluster, indicating a parameter configuration less likely to exhibit repeat activity.

While the Repeater-like cluster is dominated by repeating FRBs, it also includes some apparent non-repeaters. Conversely, the Nonrepeater-like cluster consists primarily of apparent non-repeaters but contains a few repeaters as well. Interestingly, some repeating bursts assigned to both clusters originate from the same repeating source. Those repeating bursts in the Nonrepeater-like cluster are rare and tend to exhibit broader bandwidths, narrower pulse widths, and higher energies than typical Repeater-like bursts, representing atypical repeating activity. Such cases demonstrate that repeaters can span a broad parameter space and occasionally resemble typical non-repeaters, reinforcing the view that the two clusters reflect \blue{distinct} but overlapping similarity patterns in the nine-dimensional parameter space, possibly shaped by different physical mechanisms or environments. Moreover, the distribution of repeaters within the Repeater-like cluster itself appears inhomogeneous, suggesting the presence of substructures or subclusters that merit detailed investigation with larger samples in future studies.

Compared to our analysis of the first CHIME/FRB catalog in \citet{Sun2025}, we observe that clearly separating repeaters from non-repeaters has become increasingly difficult, as reflected in the significant increase in the proportion of repeaters assigned to the Nonrepeater-like cluster, from $0.3\%$ to $4.9\%$ in this study. These atypical repeaters account for approximately $12\%$ of all repeaters. This rise indicates that parameter characteristics once thought exclusive to non-repeaters can also appear in repeating sources, pointing to an increasingly continuous distribution of FRBs in feature space. This trend is corroborated by follow-up observations \citep{2023ApJ...947...83C}, which show that repeaters such as FRB 20181226F and FRB 20200127B exhibit broader bandwidths and narrower pulse widths-features previously associated with non-repeaters. This finding implies that future investigations should explore the diversity and continuity in the parameter-space distribution of repeaters in more detail, and examine the underlying physical processes or evolutionary scenarios that could account for the increasingly blurred observational boundary between repeating and non-repeating FRBs.

\subsection{Feature Importance} \label{ssec:feature}

\begin{figure}
\centering
\includegraphics[angle=0,scale=0.62]{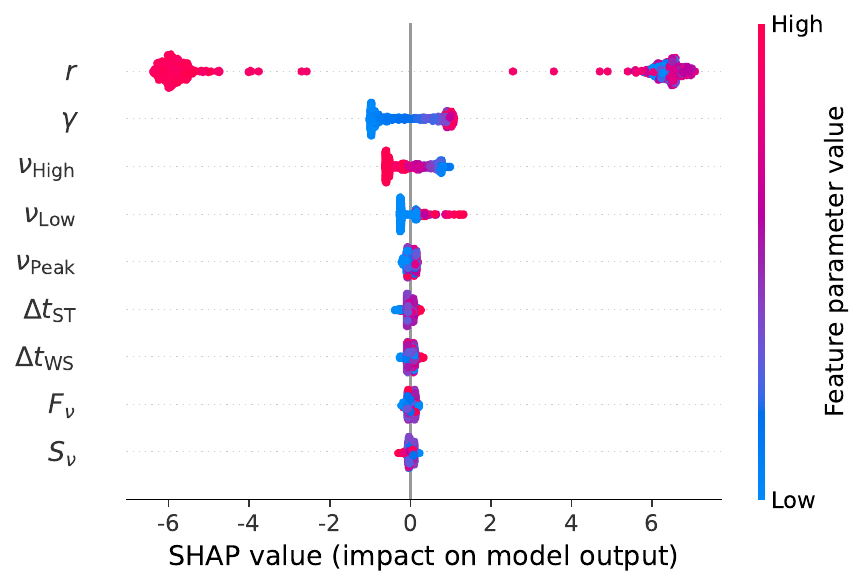}
\caption{
SHAP values for predictions in FRB classification. Features are ranked by overall importance, with the horizontal axis indicating the SHAP value, reflecting each feature’s impact on the model output.
\label{fig:figure2}}
\end{figure}

To further investigate the dominant roles of observational parameters in distinguishing the Repeater-like and Nonrepeater-like clusters, we construct a supervised classification model based on the CatBoostClassifier \citep{Prokhorenkova2018}, building on the HDBSCAN clustering results. We then apply the SHAP \citep[SHapley Additive exPlanations,][]{Lundberg2017} method to quantitatively interpret the contribution of each feature to the model output and to identify the most influential parameters. SHAP is a game-theoretic model interpretation framework that estimates the marginal contribution of each feature to the model prediction, thereby revealing both the importance and directional influence of each feature.

Figure \ref{fig:figure2} presents the distribution of SHAP values for each feature parameter. The vertical axis lists the input features, sorted from top to bottom by their relative importance in influencing the model's classification of FRBs into the two clusters. The horizontal axis shows the magnitude and direction of the SHAP values, reflecting each feature’s contribution and tendency toward classifying an event as belonging to either the Nonrepeater-like or Repeater-like cluster. The color gradient from blue to red indicates the variation in the corresponding feature values from low to high. It is important to note that, during model training, we define the Repeater-like cluster as the positive class. Accordingly, a positive SHAP value implies that the corresponding feature value pushes the model toward classifying the sample as part of the Repeater-like cluster, while a negative SHAP value indicates a tendency toward classification into the Nonrepeater-like cluster.

As shown in Figure \ref{fig:figure2}, the spectral morphology parameters, i.e., $r$ and $\gamma$, emerged as the top two most predictive features. These two features contribute significantly more than the others, consistent with our previous findings based on permutation feature importance analysis \citep{Sun2025}. 
This result indicates that the Repeater-like and Nonrepeater-like clusters exhibit pronounced differences in their spectral characteristics. Specifically, FRBs with lower (more negative) $r$ values and larger $\gamma$ values are more likely to be assigned to the Repeater-like cluster. This is aligned with the physical picture in which repeating FRBs tend to exhibit narrower spectral morphologies \citep{Ziggy2021}, potentially reflecting a preferred spectral morphology associated with their radiation mechanisms or propagation effects.

In addition, SHAP analysis highlights the significant contributions of the frequency-related parameters $\nu_\mathrm{High}$ and $\nu_\mathrm{Low}$ to cluster classification. Specifically, FRBs with higher $\nu_\mathrm{Low}$ and lower $\nu_\mathrm{High}$, indicating a narrower bandwidth, tend to be classified into the Repeater-like cluster. Conversely, FRBs with broader bandwidth (i.e., lower-frequency minima and higher-frequency maxima) are more likely to fall within the Nonrepeater-like cluster. In contrast, temporal parameters ($\Delta t_\mathrm{ST}$ and $\Delta t_\mathrm{WS}$) and intensity-related parameters ($F_{\nu}$ and $S_{\nu}$) are assigned lower importance in the SHAP ranking. Nevertheless, $\Delta t_\mathrm{ST}$ and $\Delta t_\mathrm{WS}$ still exhibit modest directional trends: FRBs in the Repeater-like cluster generally have broader pulse widths, whereas those in the Nonrepeater-like cluster tend to show narrower widths. The spectral and temporal trends identified here through SHAP analysis are broadly consistent with previous findings, such as those by \citet{2021ApJS..257...59C,Ziggy2021,2023ApJ...947...83C,2025ApJ...992..206C}.

\subsection{$\gamma$-$r$ Parameter Distributions} \label{ssec:gamma-r}
\subsubsection{$\gamma$-$r$ Space Number Density Distribution} \label{sssec:distribution gamma-r}
Based on the preceding SHAP analysis, this section focuses on the two most discriminative spectral parameters, $\gamma$ and $r$, which effectively separate the Repeater-like and Nonrepeater-like clusters. According to the definition in Equation \ref{equ:equ1}, the parameter $r$ is closely related to the emission bandwidth and effectively characterizes the spectral turnover. This formulation captures a wide range of spectral morphologies: when $r \sim 0$, the model reduces to a standard power law, indicative of broadband emission; in contrast, negative values of $r$ correspond to spectra with a well-defined peak, characteristic of narrowband emission.

\begin{figure}
\centering
\includegraphics[angle=0,scale=0.52]{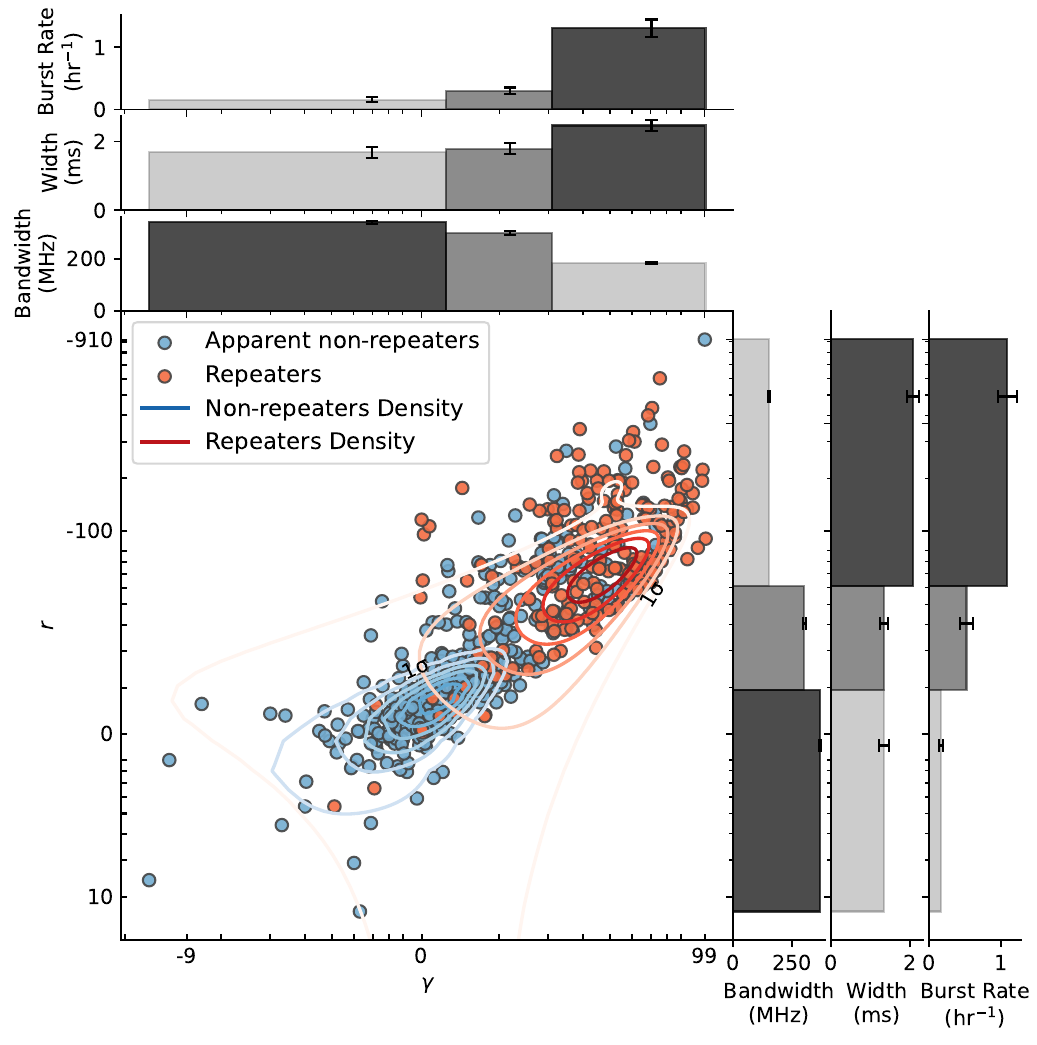}
\caption{Distribution of repeaters (red) and apparent non-repeaters (blue) in the $\gamma$-$r$ parameter space, with red and blue contours overlaid to represent the density levels of each population. Marginal histograms show the distributions of bandwidth, pulse width, and burst rate, with error bars indicating the standard errors of the means.
\label{fig:figure3}}
\end{figure}

In this section, we aim to construct a simple observational framework using only $r$ and $\gamma$ to assess their ability to systematically differentiate repeaters from apparent non-repeaters. Furthermore, we investigate their statistical correlations with the average burst rate, frequency bandwidth, and pulse width, in order to gain deeper insights into the relationship between repeating behavior and observed diversity. Figure \ref{fig:figure3} illustrates the distribution of repeating (red) and apparently non-repeating (blue) FRBs in $\gamma$-$r$ parameter space.
Although both repeaters and apparent non-repeaters are present across the entire $\gamma$-$r$ space, indicating overlap in their observed properties, they exhibit distinct preferences.
The number density distributions of repeaters and apparent non-repeaters
are respectively fitted using the Gaussian Mixture Model
\citep[GMM\footnote{\url{https://scikit-learn.org/stable/modules/mixture.html\#gaussian-mixture}};][]{2011JMLR...12.2825P}. 
The best-fit results are shown as red and blue contours in Figure \ref{fig:figure3}, 
where the outermost contour encloses 68\% of the total sample. We observe distinct peak number densities for the two populations, indicating a systematic difference between repeaters and apparent non-repeaters. Nevertheless, the 68\% region of the repeaters substantially overlaps with the density peak of the apparent non-repeaters. 
Specifically, most apparent non-repeaters, located in the high number-density region,
exhibit small $\gamma$ values and $r \sim 0$, indicating relatively flat spectral
profiles with minimal curvature, typically associated with broadband emission.
In contrast, repeaters show a broader distribution, with their density peak shifted
toward larger $\gamma$ and more negative $r$ values, and a low-density tail extending toward more extreme spectral shapes. This pattern suggests steeper spectra and narrower-band emission features.

To further evaluate the discriminative power of the two spectral parameters, $r$ and $\gamma$, we calculate their combined classification performance. The resulting AUC \citep[area under the receiver operating characteristic curve,][]{FAWCETT2006861} is 0.89, demonstrating strong separation between repeaters and non-repeaters and confirming $r$ and $\gamma$ as reliable primary discriminators.

Overall, the $\gamma$-$r$ space exhibits both continuity and multimodality. The significant separation between the density peaks of repeaters and non-repeaters suggests that repetition correlates with specific spectral properties, potentially reflecting intrinsic differences in emission or propagation mechanisms. However, current evidence remains insufficient to conclusively determine whether repeaters and non-repeaters originate from fundamentally distinct source classes. A more cautious interpretation is that the observed separation represents two characteristic regimes, possibly reflecting extremes of a continuous spectrum of behaviors produced by a single population or physical process operating under different conditions, as discussed in Section \ref{sssec:interpretation}.

\subsubsection{$\gamma$-$r$ Space Correlation with Bandwidth, Pulse Width, and Burst Rate} \label{sssec:correlation}
In the $\gamma$-$r$ parameter space, we further examine the systematic variations of frequency bandwidth, pulse width, and FRB burst rate across different regions. Both bandwidth and pulse width have long been suggested to correlate statistically with repeating behavior \citep{2016ApJ...833..177S,2019ApJ...885L..24C,2020ApJ...891L...6F,2021ApJS..257...59C,2022ApJ...926..206Z,2023ApJ...947...83C,Ziggy2021,2024ChPhL..41j9501Z}. 
Meanwhile, burst rate, defined as the number of detected bursts per unit time, represents an observational upper limit on the repeatability of an FRB over the observational timescale. The burst rate is estimated via
\begin{equation}
{\rm Burst\ Rate} = \frac{N}{t_{\rm expo}}\left(\frac{F_{0}}{F_{\rm th}}\right)^{-1.5},
\end{equation}
where $N$ is the number of detected bursts, $t_{\rm expo}$ is the exposure time, 
$F_{0}=5\ {\rm Jy \cdot ms}$ is the adopted reference fluence threshold, 
and $F_{\rm th}$ denotes the observational fluence threshold for each FRB,
provided by the CHIME/FRB collaboration. 
For the specific computation, we follow the method outlined by \citet{2023ApJ...947...83C}, while for a more comprehensive statistical analysis, we include bursts detected during both
the upper transit and lower transit.
The exposure time is determined based on the precise sky position of each source \citep{2019ApJ...885L..24C}. 
We reconsider the exposure time correction with both the transit status
and the frequency-dependent beam width for each burst. 
To ensure uniform sensitivity across different sources, we adopt the fluence thresholds provided by the CHIME/FRB collaboration \citep{2021ApJS..257...59C,2023ApJ...947...83C}
using the method of \citet{2019ApJ...882L..18J}, which accounts for variations
of system gain, beam response, and bandwidth, and applies the 95th percentile
of the scaled threshold distribution per source. 
Finally, the burst rates for all sources are normalized to a fluence 
threshold of $5\,{\rm Jy \cdot ms}$, assuming a power-law energy distribution
with an index of $-1.5$ \citet{2023ApJ...947...83C}.

In our analysis, we divide the sample into three equally populated bins along both $r$ and $\gamma$ directions to ensure statistical robustness. Within each bin, we calculate the average frequency bandwidth, pulse width, and burst rate. The bar plots in Figure \ref{fig:figure3} visualize these statistics, with color gradients indicating the variation in average $\gamma$ and $r$ values. The results reveal a consistent trend: as one moves from the density peak of non-repeaters toward that of repeaters in the $\gamma$-$r$ parameter space, the frequency bandwidth gradually decreases, the pulse width increases, and the burst rate rises significantly. 
Our statistical analysis thus reveals both the positive correlation between repetition rate and pulse width and the negative correlation with bandwidth, consistent with the predictions of \citet{2023ApJ...947...83C} and supporting their proposed continuum framework that links burst morphology to repetition rate in a single repeating FRB population. This monotonic trend from the non-repeaters to the repeaters region further reinforces the link between spectral morphology and repeatability, supporting the scenario that spectral morphology serves as an important diagnostic for understanding the physical origin of FRBs.

\subsubsection{Physical Interpretation within Geometric Frameworks} \label{sssec:interpretation}
{Based on the above analysis, FRBs in the $\gamma$-$r$ space exhibit a positive correlation between repetition rate and pulse width and a negative correlation with bandwidth. These trends define a continuous, monotonic transition from non-repeaters to repeaters, suggesting that the two populations may not arise from fundamentally distinct progenitor mechanisms.} Instead, they more likely represent two manifestations of the same underlying mechanism or source class under different geometric and physical conditions, i.e., two extreme cases within a continuous distribution. The rotating polar cap scenario \citep{2025ApJ...982...45B,2025ApJ...988...62L} provides a coherent framework for interpreting these results.
In this framework, whether an FRB appears as repeating or non-repeating, along with its associated spectral and temporal characteristics, is mainly governed by the viewing geometry of the line of sight relative to the magnetar’s emission cone. In the aligned-rotator geometry ($\alpha \ll \rho$), where $\alpha$ is the magnetic inclination angle and $\rho$ is the half-opening angle of the emitting beam in the observer’s frame, the line of sight remains within a fixed emission region, so the physical conditions vary little, leading to pulse widths set mainly by the intrinsic burst timescale and spectra centered around a stable frequency. Such emission conditions can be naturally sustained in magnetars residing in binary systems \citep{2025arXiv250812119Z}. This produces bursts with narrower spectra, longer widths, and higher repetition rates, corresponding to the clustering of repeating FRBs at $r \ll 0$ and relatively large $\gamma$ in Figure \ref{fig:figure3}. 
In contrast, in the misaligned-rotator geometry ($\alpha \gg \rho$), the line of sight samples a wide range of surface regions as the beam sweeps by, causing large variations in emission conditions and yielding bursts with sweep-limited widths, broadened spectra, and lower apparent repetition rates. These correspond to the clustering of non-repeating FRBs around $r \sim 0$ and relatively smaller $\gamma$ in the $\gamma$-$r$ space. This geometric framework naturally explains the pronounced separation yet partial overlap of the density centers of repeating and non-repeating FRBs observed in the $\gamma$-$r$ plane {(see Figure \ref{fig:figure3}).}

The geometry framework offers a potential explanation for the observed differences between repeating and apparently non-repeating FRBs, although doing so hinges on restrictive geometric parameters. In particular, to simultaneously reproduce the nearly flat polarization position angles (PAs) commonly observed in both repeating and apparently non-repeating FRBs \citep{2018Natur.553..182M, 2019ApJ...885L..24C, 2020Natur.586..693L, 2021ApJ...908L..10H, 2021MNRAS.508.5354H, 2022Natur.611E..12X, 2022RAA....22l4003J, 2022ApJ...932...98S, 2022NatAs...6..393N, 2023ApJ...955..142Z, 2023ApJ...950...12M, 2023MNRAS.526.3652K, 2024ApJ...968...50P, 2024ApJ...974..274F, 2025ApJ...988..175L, 2025ApJS..278...49X, 2025ApJ...982..154N}, the rotating vector model \citep{1969ApL.....3..225R} requires that at least one of the following conditions be satisfied:

(1) an exceptionally long rotational period $P \gg \Delta t_\mathrm{WS}$;

(2) nearly aligned rotational and magnetic axes $\sin(\alpha) \sim 0$;

(3) a very small impact angle $|\beta| \sim 0$.

For condition (1), the requirement of a long rotational period is expected for repeating FRBs, as current statistical analyses of large samples from some hyperactive repeating FRBs have excluded the existence of periods in the millisecond-to-second range \citep{2021Natur.598..267L, 2022Natur.611E..12X, 2022RAA....22l4004N, 2023ApJ...955..142Z, 2025arXiv250715790W, 2025arXiv250714708Z, Zhang:2025qzn}.
In contrast, such a long period is not anticipated for non-repeating FRBs, since their pulse width $\Delta t_\mathrm{WS}$ relies on the rotational cut-off. For condition (2), near alignment between the spin and magnetic axes is also plausible for repeating FRBs. However, for non-repeating FRBs this condition further requires $\alpha \gg \rho$. Under this premise, both $\alpha$ and $\rho$ must be very small,  which constitutes a highly restrictive constraint. For condition (3), the requirement of a very small impact angle applies equally to both types of FRBs. Beyond the geometric conditions discussed above, an alternative explanation for the observed disparities between the repeating and non-repeating FRBs may lie in intrinsic variations within the emission mechanism itself, which also governs the resulting spectral morphology.


In addition, the power-law model for the frequency drift rate proposed by \citet{2022ApJ...925..135M} naturally aligns with this geometric interpretation and provides a physical basis for the distribution differences between repeating and non-repeating FRBs observed in Figure \ref{fig:figure3}. This model characterizes the time-frequency evolution of FRBs by describing the drift rate of the central emission frequency using a power-law index $\bar{\beta}$, such that $\nu_c(t) \propto t^{-\bar{\beta}}$: smaller values of $\bar{\beta}$ correspond to more gradual frequency drifts, resulting in narrower spectra and longer widths, whereas larger $\bar{\beta}$ values lead to rapid frequency drifts, producing broader spectra and shorter widths. Combining this with our results, the parameters $r$ and $\gamma$ can be understood as observational descriptors of such time-frequency evolution: repeating FRBs cluster in the region with $r \ll 0$ and relatively large $\gamma$, consistent with the small-$\bar{\beta}$ scenario, whereas non-repeating FRBs are more distributed around $r \sim 0$ with smaller $\gamma$, corresponding to larger $\bar{\beta}$.

We note that in Figure \ref{fig:figure3}, a small subset of repeaters falls within the primary density peak of apparent non-repeaters. These bursts exhibit broader spectra and shorter durations, and generally originate from only a few sporadic events of a given repeater. Their properties differ markedly from those of the majority of typical repeating bursts. This suggests that the emission region or emission mechanism of repeaters may not always remain stable and consistent. Instead, different physical processes, such as changes in the emission site or beaming angle, may occasionally produce short-duration, broadband, and potentially high-energy bursts that resemble those of apparently non-repeating sources \citep{2024NatAs...8..337K}. In addition, such “outlier” events may also be explained by low-probability processes external to the emission source. For example, propagation effects such as plasma lensing can, on rare occasions, significantly modify the observed spectra of bursts \citep{2017ApJ...842...35C,2021MNRAS.500..272S,2024ApJ...974..160K}. However, their occurrence rate is expected to be extremely low, and thus they can only account for a small number of anomalous events. On the other hand, statistical fluctuations in a limited sample can naturally produce a few events located at the tails of the underlying parameter distribution, making these outliers statistically plausible. Consequently, the continuous distribution and density structure differences observed in the $\gamma$-$r$ space may not only arise from geometric effects within a single theoretical framework, but also reflect continuous manifestations of the same class of sources in different physical states \citep{Beniamini2025}. This provides important observational clues for probing the origins and emission mechanisms of FRBs.

\subsection{$\gamma$-$r$ Framework for Repetition Probability Estimation} \label{ssec:probesti}
\begin{figure}
\centering
\includegraphics[angle=0,scale=0.52]{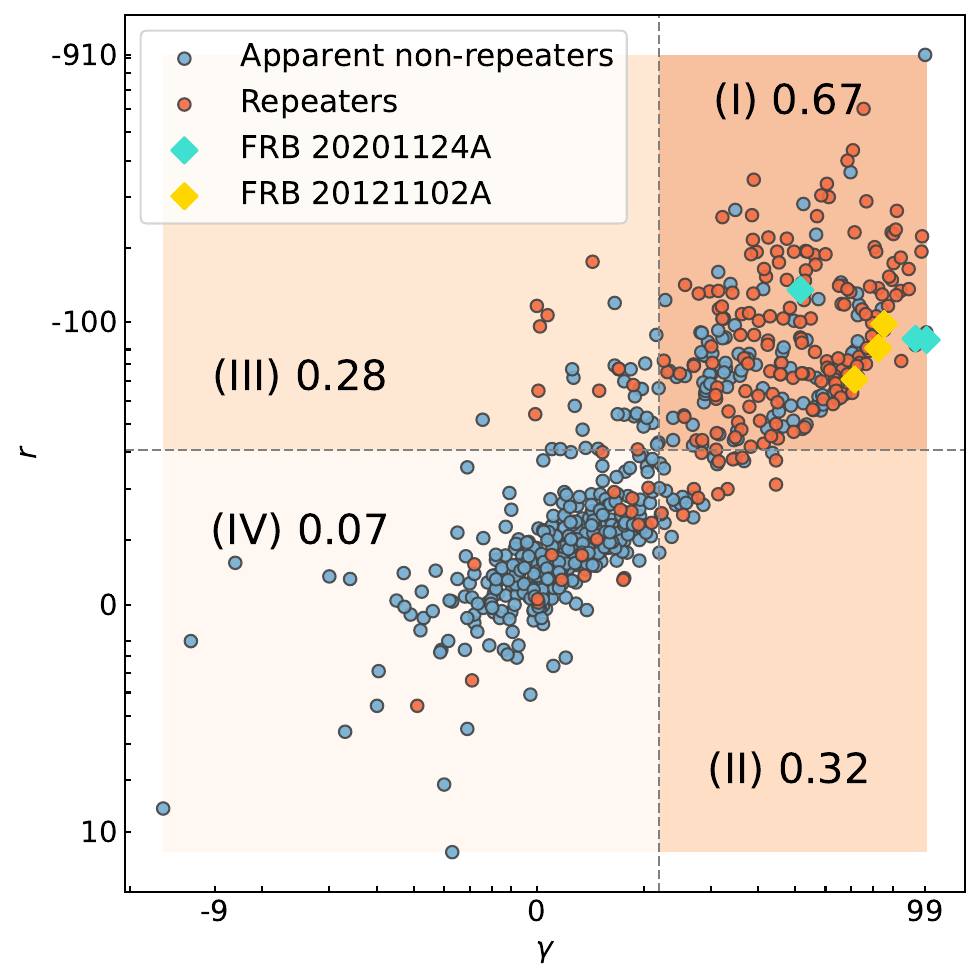}
\caption{Distribution of repeaters (red) and apparent non-repeaters (blue) in the $\gamma$-$r$ parameter space. The background shading indicates regions of different inferred repetition probabilities, with annotated values representing the mean posterior probabilities derived from the Bayesian GMM analysis. FRB 20201124A and FRB 20121102A are highlighted as cyan and yellow diamonds, respectively.
\label{fig:figure4}}
\end{figure}

In this section, we construct an empirical probability map in the $\gamma$-$r$ space to estimate the likelihood of FRBs exhibiting repeating behavior. In Figure \ref{fig:figure4}, we draw dashed lines at the median values of $r$ and $\gamma$ for the entire FRBs, dividing the parameter space into four quadrants. This division approximately corresponds to the natural boundary between the density peaks of non-repeaters and repeaters shown in Figure \ref{fig:figure3}, reflecting that certain spectral morphologies are more likely to produce repeating FRBs.

To further quantify this distinction, we incorporated a Bayesian posterior probability assessment based on GMM fitting to model the probability density of repeaters and non-repeaters in the $\gamma$-$r$ parameter space, thereby estimating the posterior probability that each FRB belongs to the repeater population. In this framework, the posterior probability for a given FRB point 
$\mathbf{x}=(\gamma, r)$ to be classified as a repeater is computed using Bayes’ theorem,  
\begin{equation}
{\mathcal{L}}(\mathrm{R} \mid \mathbf{x}) =
\frac{p(\mathbf{x} \mid \mathrm{R})\,\pi_{\mathrm{R}}}{
p(\mathbf{x} \mid \mathrm{R})\,\pi_{\mathrm{R}} + p(\mathbf{x} \mid \mathrm{N})\,\pi_{\mathrm{N}}},
\end{equation}
where $p(\mathbf{x} \mid \rm R)$ and $p(\mathbf{x} \mid \rm N)$ denote the
class likelihood estimated by fitting the GMM separately to the repeater
and apparent non-repeater number density distribution (as shown with the contours in
Figure~\ref{fig:figure3}),
and $\pi_{\rm R}$ and $\pi_{\rm N}$ represent the corresponding 
prior probabilities, taken as the observed fractions of repeaters and non-repeaters,
\begin{equation}
\pi_{\rm R} = 1 - \pi_{\rm N}.
\end{equation}
This formulation enables a probabilistic classification in which each FRB is assigned a continuous posterior probability of belonging to the repeater population, rather than a hard binary label.

Based on these posterior probabilities, we computed the average repetition probability within each region defined in Figure \ref{fig:figure4}, obtaining values of 0.67 (I), 0.32 (II), 0.28 (III), and 0.07 (IV). This pronounced gradient demonstrates a clear dependence of FRB repeatability on the $\gamma$-$r$ parameter space. In particular, the stark contrast between the high-repetition Region I and the nearly non-repeating Region IV further quantifies the association between spectral morphology and repeat behavior. This result is consistent with the previous analysis based on density contours, reinforcing spectral morphology as a potential indicator for distinguishing repeaters.

Additionally, in Figure \ref{fig:figure4}, we highlight two well-studied and highly active repeating FRB sources: FRB 20121102A and FRB 20201124A. Both have been monitored extensively with high sensitivity by the Arecibo telescope \citep{2023MNRAS.519..666J} and FAST \citep{2021Natur.598..267L,2022Natur.609..685X}, yielding records of thousands of bursts, thereby providing critical insights into the origins of FRBs. Notably, both sources fall within the high-repetition-probability Region I defined in our framework, serving as compelling evidence for the robustness and predictive usefulness of this spectral partitioning. Except for a small number of atypical repeating bursts, this framework successfully {assigns approximately 80\% of known} repeaters to the high-probability Region I. Overall, this probability map directly links FRB repeatability to spectral morphology parameters, offering a simple, physically interpretable prediction framework independent of prior labeling.

To assess the performance of the $\gamma$-$r$ framework in assigning repetition probability scores, we randomly withhold 10\% of known repeaters (corresponding to 21 FRBs) and 10\% of apparently non-repeating FRBs (49 FRBs) from the construction of the $\gamma$-$r$ framework, and build the framework using the remaining FRB sample. We then estimate the repetition probability scores for the held-out known repeaters.

\begin{figure*}
\centering
\includegraphics[angle=0,scale=0.52]{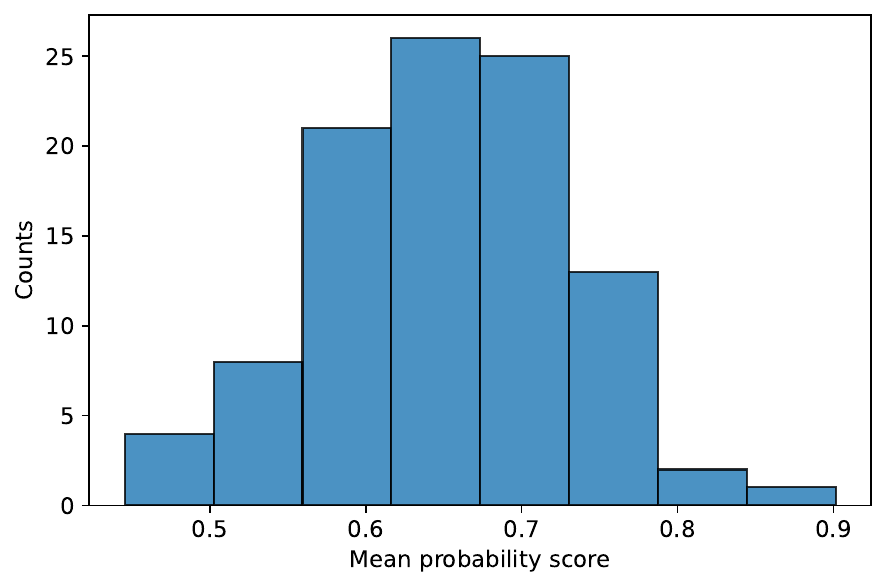}
\includegraphics[angle=0,scale=0.52]{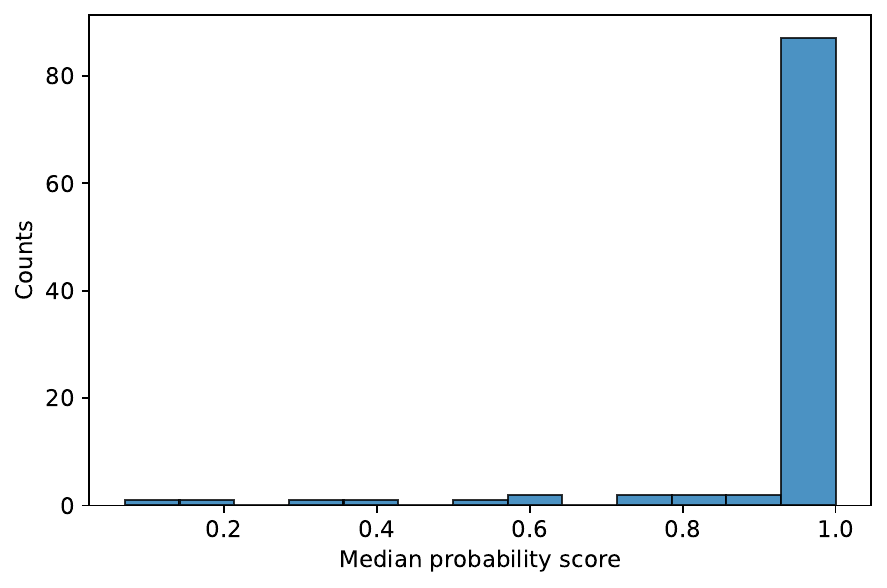}
\caption{{The left and right panels show the distributions of the mean and median repetition probability scores, respectively, for the randomly held-out 10\% known repeaters, obtained from 100 random subsampling realizations, with repetition probability scores estimated using the $\gamma$-$r$ framework.}}
\label{fig:figure5}
\end{figure*}

Given the limited number of confirmed repeating FRBs in the current dataset, we repeat this procedure 100 times to evaluate the robustness of the framework under different random realizations. Figure \ref{fig:figure5} shows the distributions of the mean (left) and median (right) repetition probability scores for the held-out repeaters across 100 random subsampling realizations. While variations in the randomly selected subsets introduce modest fluctuations in the inferred scores, the mean repetition probability score averaged over all realizations is 0.65, generally consistent with the repetition probability density of Region I in the full $\gamma$-$r$ framework (Figure~\ref{fig:figure4}). 

The median scores, which characterize the typical repetition probability of the held-out repeaters and are insensitive to a small number of extreme values, are strongly concentrated in the range 0.9-1.0 for the vast majority of realizations, with only a few cases extending into the intermediate-probability regime, consistent with the presence of a small number of repeaters with atypical spectral properties. This behavior indicates that, even when the framework is constructed from different data subsets, at least half of the held-out known repeaters are consistently assigned high repetition probabilities, demonstrating the robustness of the $\gamma$-$r$ framework.

For future FRBs newly detected by CHIME, their positions in this map can be readily determined using only $r$ and $\gamma$, enabling a preliminary estimation of their likelihood to exhibit repeating FRBs. This provides a practical and accessible tool for rapidly assessing repetition probability and prioritizing monitoring targets in future observations. Compared with machine learning classification methods that rely on multi-parameter dimensionality reduction and clustering, this physically based partitioning is faster, more intuitive, and yields robust, generalizable results.

We further combine the t-SNE results with the $\gamma$-$r$ parameter map to identify a set of FRB sources that have not yet been observed to repeat but exhibit statistical characteristics closely resembling those of known repeaters. Specifically, we first selected all non-repeating FRBs classified within the Repeater-like cluster and categorized them according to their quadrant positions in the $\gamma$-$r$ parameter space, with particular attention to those located in the high-repetition-probability Region I.

We propose that these sources, which exhibit a high repetition tendency under multiple criteria, should be prioritized as high-probability candidates for follow-up monitoring with future telescopes. Appendix Table \ref{tab:tab1} lists the detailed information for these candidates, including their quadrant positions in the $\gamma$-$r$ space, posterior probability scores derived from the GMM modeling, and corresponding $r$ and $\gamma$ values, providing a reference for future searches for repeaters and studies of spectral evolution.

\section{Conclusion} \label{sec:concl}
In this work, we analyzed the repeatability characteristics of FRBs using an extended CHIME/FRB sample, integrating dimensionality reduction, clustering, feature-importance evaluation, and spectral-parameter-based partitioning. We further propose a simple and generalizable method that relies solely on spectral parameters to classify FRBs by their repeatability preference and to effectively estimate their repetition probabilities. The main conclusions are as follows:
\begin{enumerate}
\item Using t-SNE for dimensionality reduction and HDBSCAN for clustering, we unsupervisedly divided the combined CHIME Catalog 1 and Catalog 2023 samples into two clusters. One cluster, designated as the Repeater-like cluster, exhibits a higher tendency to repeat and contains the majority of known repeaters, showing statistically significant differences from the other cluster, termed the Nonrepeater-like cluster. SHAP-based feature-importance analysis confirms that the spectral running $r$ and spectral index $\gamma$ are the dominant parameters separating repeaters from apparent non-repeaters, reinforcing the connection between spectral features and FRB repeatability. Consistently, we find that using $r$ and $\gamma$ alone yields an AUC of 0.89, demonstrating strong discriminative power and validating their role as primary distinguishing metrics.
\item In the $\gamma$-$r$ space, we find that the density centers of repeaters and non-repeaters are clearly separated, though substantial overlap exists across the entire space. Furthermore, we identify a positive correlation between burst rate and pulse width, and a negative correlation between burst rate and bandwidth in the $\gamma$-$r$ space. These trends not only reinforce the physical link between spectral morphology and repeatability but also provide key insights supporting the possibility that FRBs belong to a single underlying population.
\item The rotating polar cap scenario and the power-law model for frequency drift rates provide a coherent explanation for the distinct clustering of repeaters and non-repeaters in the $\gamma$-$r$ space, linking these distributions to geometric configurations and resulting variations in emission-region conditions.
\item We construct a four-quadrant probability map in the $\gamma$-$r$ space to rapidly assess the likelihood of repetition for different spectral parameter combinations. This map exhibits a pronounced statistical gradient, with repetition likelihoods of $\sim$67\% in the high-repetition region and only $\sim$7\% in the low-repetition region, demonstrating that the spectral morphology can serve as an effective predictor of repeatability tendencies. Notably, all three bursts from FRB 20201124A fall within the high-probability Region I, validating the robustness of this framework. Thus, it provides a simple, quantifiable, and easily generalizable tool for predicting repetition probabilities, serving as a practical reference for future telescope observations and studies of FRB repeatability. 
In addition, we identify 40 apparent non-repeaters with probability scores above 0.6 in Region I and recommend them as priority targets for future monitoring (details in the Appendix).
\end{enumerate}

In summary, this study reveals a significant statistical association between FRB repeatability and their spectral parameters, demonstrating that repetition probabilities can be effectively predicted using only simple spectral-shape parameters. This approach establishes a paradigm for a label-independent, physically motivated probabilistic framework, providing practical guidance for the rapid identification of repeaters and the design of follow-up observation strategies. 

\section*{Acknowledgments}
We are very grateful to Alice P. Curtin, Ziggy Pleunis, and Mary Jiang for helpful discussions. We acknowledge the support of the National Natural Science Foundation of China (Grants Nos. 12473001, 12533001, and 12575049), the National SKA Program of China (Grant Nos. 2022SKA0110200, 2022SKA0110203), the China Manned Space Program (Grant No. CMS-CSST-2025-A02), and the 111 Project (Grant No. B16009). F.-W.Z. acknowledges the supports from the National Natural Science Foundation of China (No. 12463008) and the Guangxi Natural Science Foundation (No. 2022GXNSFDA035083). Y.-K.Z. is supported by the Postdoctoral Fellowship Program and China Postdoctoral Science Foundation (Grant Number BX20250158).

\section*{Data Availability}
The list of apparent non-repeaters identified in this work as high-repetition-potential candidates is provided in Table \ref{tab:tab1} in APPENDIX. All necessary data used in this analysis are included in this article and the Appendix.

\bibliography{FRB_tSNE}

\section*{APPENDIX}
\label{app:appendix}
This work uses t-SNE embedding and clustering analysis based on multi-dimensional observational parameters, combined with the definition of high-repetition-probability regions in the $\gamma$-$r$ parameter space, to identify a set of FRB sources that have not yet been observed as repeaters but lie significantly close to the known repeater distribution in feature space. Specifically, we selected all apparent non-repeaters assigned to the Repeater-like cluster, with special emphasis on those falling within the high-repetition-probability Region I in the $\gamma$-$r$ space, designating them as the highest-priority targets for future monitoring campaigns.

Table \ref{tab:tab1} provides detailed information for these candidate sources, including their Right Ascension (R.A.), Declination (Decl.), Dispersion Measure (DM), and the nine feature parameters used in our analysis, i.e., $S_{\nu}$, $F_{\nu}$, $\Delta t_\mathrm{WS}$, $\Delta t_\mathrm{ST}$, $\gamma$, $r$, $\nu_\mathrm{Low}$, $\nu_\mathrm{High}$, and $\nu_\mathrm{Peak}$. Table \ref{tab:tab1} also indicates the assigned Region for each FRB in the $\gamma$-$r$ parameter space.

\startlongtable
\begin{deluxetable*}{lccccccccccccccc}
\setlength{\tabcolsep}{3pt}
\tablecaption{High-Repetition-Potential FRB Candidates Identified from Apparent Non-repeaters. \label{tab:tab1}}
\tablewidth{700pt}
\tabletypesize{\scriptsize}
\tablehead{
\colhead{Name} & \colhead{Sub\_num} & \colhead{R.A. (J2000)} & \colhead{Decl. (J2000)} & \colhead{DM}& \colhead{$S_{\nu}$\vspace{-2pt}} & \colhead{$F_{\nu}$\vspace{-2pt}} & \colhead{$\Delta t_\mathrm{WS}$} & \colhead{$\Delta t_\mathrm{ST}$} & \colhead{$\gamma$} & \colhead{$r$} & \colhead{$\nu_\mathrm{Low}$\vspace{-2pt}} & \colhead{$\nu_\mathrm{High}$\vspace{-2pt}} & \colhead{$\nu_\mathrm{Peak}$\vspace{-2pt}} & \colhead{Quadrant} & \colhead{Probability Score}  \\
\colhead{} & \colhead{} & \colhead{(\(^\circ\)\vspace{-2pt})} & \colhead{(\(^\circ\)\vspace{-2pt})}  & \colhead{(pc cm$^{-3}$)}& \colhead{(Jy)} & \colhead{(Jy ms)} & \colhead{(ms)} & \colhead{(ms)} & \colhead{} & \colhead{} & \colhead{(MHz)} & \colhead{(MHz)} & \colhead{(MHz)} & \colhead{} & \colhead{}  \\[-14pt]
}

\startdata
FRB20190423B	&	0	&	$298.58^{+0.210}_{-0.210}$	&	$26.19^{+0.210}_{-0.210}$	&	584.949 	&	0.87 	&	7.00 	&	2.490 	&	3.000 	&	$	62.40 	$	&	$	-106.00 	$	&	463.80 	&	623.10 	&	537.60 	&	I	&	0.95 	\\
FRB20181017B	&	0	&	$237.76^{+0.230}_{-0.230}$	&	$78.50^{+0.250}_{-0.250}$	&	307.369 	&	1.06 	&	6.50 	&	2.310 	&	4.300 	&	$	61.00 	$	&	$	-77.00 	$	&	499.30 	&	704.80 	&	593.20 	&	I	&	0.95 	\\
FRB20190422A	&	1	&	$48.56^{+0.200}_{-0.200}$	&	$35.15^{+0.200}_{-0.200}$	&	452.302 	&	0.60 	&	9.10 	&	2.310 	&	2.700 	&	$	54.20 	$	&	$	-63.70 	$	&	506.30 	&	740.50 	&	612.30 	&	I	&	0.95 	\\
FRB20190609A	&	1	&	$345.30^{+0.280}_{-0.280}$	&	$87.94^{+0.330}_{-0.330}$	&	316.643 	&	3.60 	&	10.40 	&	2.120 	&	0.500 	&	$	55.00 	$	&	$	-67.00 	$	&	499.10 	&	722.40 	&	600.50 	&	I	&	0.95 	\\
FRB20190112A	&	0	&	$257.98^{+0.015}_{-0.015}$	&	$61.20^{+0.026}_{-0.026}$	&	425.847 	&	1.40 	&	16.20 	&	1.640 	&	11.010 	&	$	57.10 	$	&	$	-51.40 	$	&	564.60 	&	800.20 	&	697.70 	&	I	&	0.95 	\\
FRB20180801A	&	0	&	$322.53^{+0.059}_{-0.059}$	&	$72.72^{+0.220}_{-0.220}$	&	655.728 	&	1.11 	&	7.90 	&	0.580 	&	5.540 	&	$	60.00 	$	&	$	-75.50 	$	&	500.20 	&	709.30 	&	595.60 	&	I	&	0.95 	\\
FRB20190623B	&	0	&	$335.22^{+0.180}_{-0.180}$	&	$46.12^{+0.180}_{-0.180}$	&	1556.765 	&	1.58 	&	2.78 	&	0.440 	&	0.720 	&	$	54.20 	$	&	$	-57.10 	$	&	526.20 	&	786.40 	&	643.30 	&	I	&	0.95 	\\
FRB20190609A	&	0	&	$345.30^{+0.280}_{-0.280}$	&	$87.94^{+0.330}_{-0.330}$	&	316.643 	&	3.60 	&	10.40 	&	0.432 	&	0.500 	&	$	62.40 	$	&	$	-84.00 	$	&	491.00 	&	683.40 	&	579.30 	&	I	&	0.95 	\\
FRB20190428A	&	0	&	$170.73^{+0.027}_{-0.027}$	&	$23.33^{+0.150}_{-0.150}$	&	969.400 	&	2.22 	&	7.40 	&	0.374 	&	3.630 	&	$	54.00 	$	&	$	-48.80 	$	&	560.10 	&	800.20 	&	696.00 	&	I	&	0.95 	\\
FRB20181129B	&	0	&	$307.56^{+0.260}_{-0.260}$	&	$81.32^{+0.340}_{-0.340}$	&	405.905 	&	4.00 	&	9.50 	&	0.364 	&	0.830 	&	$	73.80 	$	&	$	-112.00 	$	&	482.30 	&	642.70 	&	556.80 	&	I	&	0.95 	\\
FRB20181231B	&	0	&	$128.77^{+0.200}_{-0.200}$	&	$55.99^{+0.180}_{-0.180}$	&	197.170 	&	0.89 	&	2.34 	&	0.337 	&	1.750 	&	$	59.60 	$	&	$	-60.00 	$	&	540.60 	&	800.00 	&	657.70 	&	I	&	0.95 	\\
FRB20181228B	&	0	&	$250.43^{+0.210}_{-0.210}$	&	$63.85^{+0.210}_{-0.210}$	&	568.651 	&	0.40 	&	1.67 	&	0.100 	&	1.159 	&	$	59.30 	$	&	$	-353.00 	$	&	401.50 	&	471.80 	&	435.20 	&	I	&	0.95 	\\
FRB20190423B	&	1	&	$298.58^{+0.210}_{-0.210}$	&	$26.19^{+0.210}_{-0.210}$	&	584.949 	&	0.87 	&	7.00 	&	8.500 	&	3.000 	&	$	63.00 	$	&	$	-116.00 	$	&	455.60 	&	604.10 	&	524.60 	&	I	&	0.90 	\\
FRB20180916C	&	0	&	$107.15^{+0.230}_{-0.230}$	&	$45.08^{+0.240}_{-0.240}$	&	2252.873 	&	0.39 	&	2.10 	&	4.060 	&	5.100 	&	$	47.00 	$	&	$	-50.00 	$	&	516.90 	&	794.60 	&	640.90 	&	I	&	0.90 	\\
FRB20190228A	&	0	&	$183.48^{+0.005}_{-0.005}$	&	$22.90^{+0.120}_{-0.120}$	&	419.083 	&	1.79 	&	35.80 	&	2.250 	&	18.910 	&	$	52.60 	$	&	$	-51.90 	$	&	538.40 	&	800.20 	&	664.70 	&	I	&	0.90 	\\
FRB20190605D	&	0	&	$26.72^{+0.230}_{-0.230}$	&	$28.62^{+0.250}_{-0.250}$	&	1656.533 	&	0.82 	&	2.16 	&	1.069 	&	1.200 	&	$	48.30 	$	&	$	-50.80 	$	&	520.40 	&	796.60 	&	643.90 	&	I	&	0.90 	\\
FRB20190329A	&	0	&	$65.54^{+0.190}_{-0.190}$	&	$73.63^{+0.270}_{-0.270}$	&	188.606 	&	0.52 	&	2.24 	&	1.040 	&	0.900 	&	$	42.00 	$	&	$	-272.00 	$	&	400.20 	&	473.90 	&	432.30 	&	I	&	0.90 	\\
FRB20181213B	&	0	&	$183.52^{+0.220}_{-0.220}$	&	$53.70^{+0.230}_{-0.230}$	&	626.593 	&	0.75 	&	1.70 	&	0.850 	&	1.200 	&	$	45.60 	$	&	$	-45.10 	$	&	529.60 	&	800.20 	&	664.00 	&	I	&	0.90 	\\
FRB20181221A	&	0	&	$230.58^{+0.200}_{-0.200}$	&	$25.86^{+0.210}_{-0.210}$	&	316.237 	&	1.25 	&	5.80 	&	0.754 	&	1.323 	&	$	62.10 	$	&	$	-128.00 	$	&	446.10 	&	583.30 	&	510.10 	&	I	&	0.90 	\\
FRB20181203B	&	0	&	$47.31^{+0.004}_{-0.004}$	&	$24.02^{+0.120}_{-0.120}$	&	375.387 	&	1.45 	&	4.50 	&	0.578 	&	2.320 	&	$	47.80 	$	&	$	-43.60 	$	&	550.80 	&	800.20 	&	693.20 	&	I	&	0.90 	\\
FRB20190519J	&	0	&	$296.21^{+0.085}_{-0.085}$	&	$86.93^{+0.300}_{-0.300}$	&	642.759 	&	0.63 	&	1.70 	&	0.460 	&	0.501 	&	$	24.30 	$	&	$	-259.00 	$	&	400.20 	&	461.00 	&	419.50 	&	I	&	0.90 	\\
FRB20181117C	&	0	&	$53.21^{+0.170}_{-0.170}$	&	$25.73^{+0.180}_{-0.180}$	&	1773.739 	&	1.57 	&	3.00 	&	0.100 	&	2.370 	&	$	48.60 	$	&	$	-46.30 	$	&	541.20 	&	800.20 	&	676.50 	&	I	&	0.90 	\\
FRB20190422A	&	0	&	$48.56^{+0.200}_{-0.200}$	&	$35.15^{+0.200}_{-0.200}$	&	452.302 	&	0.60 	&	9.10 	&	3.220 	&	2.700 	&	$	42.00 	$	&	$	-46.90 	$	&	501.70 	&	781.40 	&	626.10 	&	I	&	0.80 	\\
FRB20190218B	&	0	&	$268.70^{+0.220}_{-0.220}$	&	$17.93^{+0.260}_{-0.260}$	&	547.868 	&	0.57 	&	5.90 	&	2.050 	&	14.100 	&	$	46.20 	$	&	$	-60.00 	$	&	483.40 	&	715.20 	&	588.00 	&	I	&	0.80 	\\
FRB20190701C	&	0	&	$96.36^{+0.230}_{-0.230}$	&	$81.63^{+0.270}_{-0.270}$	&	974.195 	&	0.88 	&	2.50 	&	1.440 	&	1.800 	&	$	46.20 	$	&	$	-211.00 	$	&	402.20 	&	495.50 	&	446.40 	&	I	&	0.80 	\\
FRB20190129A	&	0	&	$45.06^{+0.210}_{-0.210}$	&	$21.42^{+0.230}_{-0.230}$	&	484.761 	&	0.49 	&	5.00 	&	1.130 	&	10.200 	&	$	43.00 	$	&	$	-37.80 	$	&	552.80 	&	800.20 	&	707.70 	&	I	&	0.80 	\\
FRB20190211A	&	0	&	$67.06^{+0.190}_{-0.190}$	&	$68.64^{+0.200}_{-0.200}$	&	1188.256 	&	1.47 	&	5.80 	&	0.360 	&	3.290 	&	$	38.90 	$	&	$	-39.30 	$	&	515.00 	&	800.20 	&	656.00 	&	I	&	0.80 	\\
FRB20190101B	&	0	&	$307.77^{+0.230}_{-0.230}$	&	$29.89^{+0.230}_{-0.230}$	&	1323.906 	&	1.02 	&	4.40 	&	0.320 	&	5.160 	&	$	41.70 	$	&	$	-36.10 	$	&	554.30 	&	800.20 	&	713.60 	&	I	&	0.80 	\\
FRB20190130B	&	0	&	$172.11^{+0.160}_{-0.160}$	&	$16.05^{+0.072}_{-0.072}$	&	989.031 	&	0.77 	&	2.95 	&	0.265 	&	0.769 	&	$	55.40 	$	&	$	-140.80 	$	&	428.60 	&	553.60 	&	487.10 	&	I	&	0.80 	\\
FRB20190125A	&	0	&	$45.73^{+0.240}_{-0.240}$	&	$27.81^{+0.260}_{-0.260}$	&	564.701 	&	0.37 	&	2.60 	&	3.210 	&	4.100 	&	$	36.00 	$	&	$	-37.00 	$	&	510.10 	&	800.20 	&	655.50 	&	I	&	0.70 	\\
FRB20180920B	&	0	&	$191.09^{+0.230}_{-0.230}$	&	$63.52^{+0.240}_{-0.240}$	&	463.400 	&	0.35 	&	1.70 	&	2.330 	&	1.730 	&	$	12.30 	$	&	$	-121.00 	$	&	400.20 	&	483.40 	&	421.10 	&	I	&	0.70 	\\
FRB20181012B	&	0	&	$206.33^{+0.210}_{-0.210}$	&	$64.15^{+0.065}_{-0.065}$	&	715.189 	&	0.49 	&	1.44 	&	0.560 	&	0.260 	&	$	20.90 	$	&	$	-154.00 	$	&	400.20 	&	483.90 	&	428.30 	&	I	&	0.70 	\\
FRB20181214A	&	0	&	$70.00^{+0.180}_{-0.180}$	&	$43.07^{+0.180}_{-0.180}$	&	468.148 	&	0.16 	&	0.41 	&	0.533 	&	0.442 	&	$	23.30 	$	&	$	-139.00 	$	&	400.20 	&	494.60 	&	435.00 	&	I	&	0.70 	\\
FRB20190624B	&	0	&	$304.65^{+0.068}_{-0.068}$	&	$73.61^{+0.200}_{-0.200}$	&	213.922 	&	16.50 	&	20.00 	&	0.372 	&	0.400 	&	$	34.80 	$	&	$	-43.20 	$	&	475.40 	&	754.30 	&	598.80 	&	I	&	0.70 	\\
FRB20180725A	&	0	&	$93.42^{+0.039}_{-0.039}$	&	$67.07^{+0.210}_{-0.210}$	&	715.809 	&	1.70 	&	4.10 	&	0.296 	&	1.100 	&	$	38.20 	$	&	$	-45.80 	$	&	485.30 	&	760.10 	&	607.40 	&	I	&	0.70 	\\
FRB20190527A	&	0	&	$12.45^{+0.200}_{-0.200}$	&	$7.99^{+0.067}_{-0.067}$	&	584.580 	&	0.47 	&	10.10 	&	2.670 	&	5.080 	&	$	47.00 	$	&	$	-122.00 	$	&	422.40 	&	556.10 	&	484.70 	&	I	&	0.60 	\\
FRB20190527A	&	1	&	$12.45^{+0.200}_{-0.200}$	&	$7.99^{+0.067}_{-0.067}$	&	584.580 	&	0.47 	&	10.10 	&	2.470 	&	5.080 	&	$	30.70 	$	&	$	-133.00 	$	&	400.20 	&	512.20 	&	449.10 	&	I	&	0.60 	\\
FRB20190403E	&	0	&	$220.22^{+0.086}_{-0.086}$	&	$86.54^{+0.270}_{-0.270}$	&	226.198 	&	3.90 	&	76.00 	&	2.200 	&	18.200 	&	$	31.70 	$	&	$	-36.20 	$	&	482.10 	&	798.70 	&	620.60 	&	I	&	0.60 	\\
FRB20181223B	&	0	&	$174.89^{+0.220}_{-0.220}$	&	$21.59^{+0.240}_{-0.240}$	&	565.655 	&	0.68 	&	4.10 	&	1.570 	&	3.500 	&	$	33.30 	$	&	$	-41.00 	$	&	473.80 	&	761.20 	&	600.60 	&	I	&	0.60 	\\
FRB20190408A	&	0	&	$262.20^{+0.230}_{-0.230}$	&	$71.60^{+0.260}_{-0.260}$	&	863.380 	&	0.64 	&	1.51 	&	0.839 	&	1.000 	&	$	35.70 	$	&	$	-49.00 	$	&	464.10 	&	716.50 	&	576.60 	&	I	&	0.60 	\\
FRB20190429B	&	0	&	$329.93^{+0.240}_{-0.240}$	&	$3.96^{+0.330}_{-0.330}$	&	295.650 	&	0.74 	&	5.00 	&	6.380 	&	7.800 	&	$	99.00 	$	&	$	-910.00 	$	&	401.70 	&	444.10 	&	422.40 	&	I	&	$<0.60$	\\
FRB20190128C	&	0	&	$69.80^{+0.230}_{-0.230}$	&	$78.94^{+0.380}_{-0.380}$	&	310.622 	&	0.71 	&	5.90 	&	6.160 	&	7.600 	&	$	22.60 	$	&	$	-55.00 	$	&	400.60 	&	603.20 	&	491.60 	&	I	&	$<0.60$	\\
FRB20181101A	&	0	&	$21.26^{+0.011}_{-0.011}$	&	$53.88^{+0.160}_{-0.160}$	&	1472.678 	&	0.50 	&	10.70 	&	6.030 	&	10.000 	&	$	16.40 	$	&	$	-37.70 	$	&	400.20 	&	636.80 	&	497.40 	&	I	&	$<0.60$	\\
FRB20181229B	&	0	&	$238.37^{+0.230}_{-0.230}$	&	$19.78^{+0.260}_{-0.260}$	&	389.047 	&	0.42 	&	4.90 	&	3.360 	&	5.100 	&	$	22.00 	$	&	$	-103.00 	$	&	400.20 	&	517.50 	&	445.50 	&	I	&	$<0.60$	\\
FRB20190409B	&	0	&	$126.65^{+0.220}_{-0.220}$	&	$63.47^{+0.210}_{-0.210}$	&	285.633 	&	0.39 	&	6.80 	&	2.340 	&	20.900 	&	$	21.10 	$	&	$	-34.10 	$	&	420.60 	&	707.30 	&	545.50 	&	I	&	$<0.60$	\\
FRB20181128C	&	0	&	$268.77^{+0.023}_{-0.023}$	&	$49.71^{+0.200}_{-0.200}$	&	618.350 	&	0.39 	&	3.40 	&	2.300 	&	2.320 	&	$	27.40 	$	&	$	-75.00 	$	&	403.20 	&	572.10 	&	480.30 	&	I	&	$<0.60$	\\
FRB20190422A	&	2	&	$48.56^{+0.200}_{-0.200}$	&	$35.15^{+0.200}_{-0.200}$	&	452.302 	&	0.60 	&	9.10 	&	2.000 	&	2.700 	&	$	24.00 	$	&	$	-32.00 	$	&	444.80 	&	763.70 	&	582.80 	&	I	&	$<0.60$	\\
FRB20181115A	&	0	&	$142.98^{+0.039}_{-0.039}$	&	$56.40^{+0.180}_{-0.180}$	&	981.613 	&	0.44 	&	1.92 	&	1.830 	&	2.100 	&	$	19.60 	$	&	$	-62.00 	$	&	400.20 	&	568.40 	&	468.80 	&	I	&	$<0.60$	\\
FRB20190403G	&	0	&	$81.74^{+0.220}_{-0.220}$	&	$25.78^{+0.250}_{-0.250}$	&	865.311 	&	0.75 	&	1.59 	&	1.590 	&	1.900 	&	$	35.70 	$	&	$	-76.00 	$	&	425.50 	&	603.20 	&	506.60 	&	I	&	$<0.60$	\\
FRB20190531C	&	0	&	$331.14^{+0.240}_{-0.240}$	&	$43.00^{+0.240}_{-0.240}$	&	478.202 	&	0.37 	&	1.20 	&	1.450 	&	1.900 	&	$	18.30 	$	&	$	-74.00 	$	&	400.20 	&	540.60 	&	453.00 	&	I	&	$<0.60$	\\
FRB20181218A	&	0	&	$5.06^{+0.190}_{-0.190}$	&	$71.35^{+0.076}_{-0.076}$	&	1874.406 	&	0.83 	&	1.59 	&	1.390 	&	0.221 	&	$	19.00 	$	&	$	-83.60 	$	&	400.20 	&	529.30 	&	448.40 	&	I	&	$<0.60$	\\
FRB20190519F	&	0	&	$165.63^{+0.200}_{-0.200}$	&	$77.23^{+0.220}_{-0.220}$	&	797.766 	&	0.75 	&	4.00 	&	1.360 	&	1.360 	&	$	21.70 	$	&	$	-86.40 	$	&	400.20 	&	534.40 	&	453.90 	&	I	&	$<0.60$	\\
FRB20180925B	&	0	&	$145.45^{+0.210}_{-0.210}$	&	$20.99^{+0.089}_{-0.089}$	&	667.866 	&	0.76 	&	2.70 	&	1.150 	&	1.400 	&	$	15.00 	$	&	$	-41.50 	$	&	400.20 	&	606.90 	&	479.60 	&	I	&	$<0.60$	\\
FRB20190629A	&	0	&	$6.34^{+0.250}_{-0.250}$	&	$12.67^{+0.260}_{-0.260}$	&	503.779 	&	0.82 	&	3.05 	&	1.140 	&	1.700 	&	$	24.70 	$	&	$	-35.30 	$	&	440.10 	&	733.60 	&	568.20 	&	I	&	$<0.60$	\\
FRB20190425B	&	0	&	$210.12^{+0.100}_{-0.100}$	&	$88.60^{+0.210}_{-0.210}$	&	1031.724 	&	1.25 	&	3.10 	&	1.108 	&	1.300 	&	$	22.40 	$	&	$	-65.60 	$	&	400.20 	&	572.60 	&	474.80 	&	I	&	$<0.60$	\\
FRB20190529A	&	0	&	$68.06^{+0.220}_{-0.220}$	&	$40.32^{+0.230}_{-0.230}$	&	704.450 	&	0.47 	&	1.45 	&	1.040 	&	1.500 	&	$	24.20 	$	&	$	-97.00 	$	&	400.20 	&	528.90 	&	453.40 	&	I	&	$<0.60$	\\
FRB20181221B	&	0	&	$306.31^{+0.004}_{-0.004}$	&	$80.98^{+0.028}_{-0.028}$	&	1395.021 	&	0.97 	&	3.30 	&	1.037 	&	1.100 	&	$	25.30 	$	&	$	-61.20 	$	&	405.40 	&	597.60 	&	492.20 	&	I	&	$<0.60$	\\
FRB20190530A	&	0	&	$68.74^{+0.025}_{-0.025}$	&	$60.59^{+0.200}_{-0.200}$	&	555.445 	&	0.58 	&	1.69 	&	1.020 	&	1.300 	&	$	17.70 	$	&	$	-91.00 	$	&	400.20 	&	517.30 	&	441.10 	&	I	&	$<0.60$	\\
FRB20190410A	&	0	&	$263.47^{+0.230}_{-0.230}$	&	$-2.37^{+0.380}_{-0.380}$	&	284.020 	&	1.59 	&	5.80 	&	1.010 	&	1.200 	&	$	43.00 	$	&	$	-85.00 	$	&	437.40 	&	607.90 	&	515.70 	&	I	&	$<0.60$	\\
FRB20190130A	&	0	&	$25.64^{+0.240}_{-0.240}$	&	$13.16^{+0.300}_{-0.300}$	&	1367.461 	&	0.47 	&	4.40 	&	0.990 	&	3.200 	&	$	19.30 	$	&	$	-62.00 	$	&	400.20 	&	567.20 	&	467.70 	&	I	&	$<0.60$	\\
FRB20190210E	&	0	&	$313.65^{+0.280}_{-0.280}$	&	$86.67^{+0.310}_{-0.310}$	&	580.580 	&	0.69 	&	1.45 	&	0.960 	&	1.100 	&	$	13.40 	$	&	$	-40.50 	$	&	400.20 	&	599.40 	&	472.20 	&	I	&	$<0.60$	\\
FRB20190304C	&	0	&	$223.01^{+0.230}_{-0.230}$	&	$26.72^{+0.250}_{-0.250}$	&	564.991 	&	0.53 	&	1.32 	&	0.948 	&	1.100 	&	$	22.30 	$	&	$	-87.00 	$	&	400.20 	&	535.20 	&	454.90 	&	I	&	$<0.60$	\\
FRB20190102A	&	0	&	$9.26^{+0.180}_{-0.180}$	&	$26.72^{+0.057}_{-0.057}$	&	699.173 	&	1.12 	&	4.20 	&	0.824 	&	0.986 	&	$	28.90 	$	&	$	-67.80 	$	&	411.90 	&	595.50 	&	495.20 	&	I	&	$<0.60$	\\
FRB20190206A	&	0	&	$244.85^{+0.220}_{-0.220}$	&	$9.36^{+0.260}_{-0.260}$	&	188.336 	&	1.40 	&	9.10 	&	0.804 	&	2.740 	&	$	38.00 	$	&	$	-65.70 	$	&	443.20 	&	644.60 	&	534.50 	&	I	&	$<0.60$	\\
FRB20181014C	&	0	&	$117.87^{+0.220}_{-0.220}$	&	$41.59^{+0.230}_{-0.230}$	&	752.167 	&	0.57 	&	1.48 	&	0.790 	&	1.000 	&	$	18.00 	$	&	$	-30.50 	$	&	408.60 	&	707.70 	&	537.70 	&	I	&	$<0.60$	\\
FRB20190223A	&	0	&	$64.72^{+0.300}_{-0.300}$	&	$87.65^{+0.320}_{-0.320}$	&	389.237 	&	0.47 	&	1.58 	&	0.763 	&	0.870 	&	$	21.80 	$	&	$	-103.00 	$	&	400.20 	&	516.50 	&	444.80 	&	I	&	$<0.60$	\\
FRB20181127A	&	0	&	$243.80^{+0.230}_{-0.230}$	&	$25.43^{+0.250}_{-0.250}$	&	930.317 	&	0.78 	&	2.90 	&	0.740 	&	0.582 	&	$	18.50 	$	&	$	-51.30 	$	&	400.20 	&	592.30 	&	479.30 	&	I	&	$<0.60$	\\
FRB20190625D	&	0	&	$115.02^{+0.013}_{-0.013}$	&	$4.87^{+0.033}_{-0.033}$	&	717.883 	&	5.30 	&	12.10 	&	0.687 	&	0.720 	&	$	17.84 	$	&	$	-76.10 	$	&	400.20 	&	535.50 	&	450.00 	&	I	&	$<0.60$	\\
FRB20190601C	&	0	&	$88.52^{+0.180}_{-0.180}$	&	$28.47^{+0.057}_{-0.057}$	&	424.066 	&	1.32 	&	5.80 	&	0.684 	&	0.119 	&	$	35.30 	$	&	$	-68.80 	$	&	430.60 	&	620.80 	&	517.00 	&	I	&	$<0.60$	\\
FRB20190518G	&	0	&	$94.79^{+0.200}_{-0.200}$	&	$75.52^{+0.140}_{-0.140}$	&	524.946 	&	0.99 	&	1.76 	&	0.666 	&	0.720 	&	$	19.80 	$	&	$	-75.30 	$	&	400.20 	&	543.60 	&	456.40 	&	I	&	$<0.60$	\\
FRB20190205A	&	0	&	$342.22^{+0.250}_{-0.250}$	&	$83.37^{+0.300}_{-0.300}$	&	695.389 	&	0.74 	&	1.70 	&	0.602 	&	0.690 	&	$	18.30 	$	&	$	-47.30 	$	&	400.20 	&	605.60 	&	485.70 	&	I	&	$<0.60$	\\
FRB20190309A	&	0	&	$278.96^{+0.230}_{-0.230}$	&	$52.41^{+0.240}_{-0.240}$	&	356.900 	&	0.39 	&	0.72 	&	0.581 	&	0.750 	&	$	12.90 	$	&	$	-64.00 	$	&	400.20 	&	535.80 	&	442.90 	&	I	&	$<0.60$	\\
FRB20181123A	&	0	&	$300.76^{+0.023}_{-0.023}$	&	$55.87^{+0.180}_{-0.180}$	&	798.718 	&	0.99 	&	2.50 	&	0.580 	&	1.920 	&	$	25.60 	$	&	$	-64.30 	$	&	404.20 	&	590.20 	&	488.50 	&	I	&	$<0.60$	\\
FRB20190106B	&	0	&	$335.63^{+0.180}_{-0.180}$	&	$46.13^{+0.180}_{-0.180}$	&	316.594 	&	1.70 	&	3.80 	&	0.578 	&	0.600 	&	$	16.58 	$	&	$	-68.00 	$	&	400.20 	&	543.50 	&	452.10 	&	I	&	$<0.60$	\\
FRB20190203A	&	0	&	$133.68^{+0.170}_{-0.170}$	&	$70.82^{+0.190}_{-0.190}$	&	420.573 	&	1.21 	&	4.00 	&	0.550 	&	0.832 	&	$	25.00 	$	&	$	-75.00 	$	&	400.20 	&	563.40 	&	472.90 	&	I	&	$<0.60$	\\
FRB20190308C	&	1	&	$188.36^{+0.026}_{-0.026}$	&	$44.39^{+0.170}_{-0.170}$	&	500.519 	&	0.47 	&	4.80 	&	0.550 	&	2.290 	&	$	13.90 	$	&	$	-60.50 	$	&	400.20 	&	545.70 	&	449.00 	&	I	&	$<0.60$	\\
FRB20190415C	&	0	&	$74.81^{+0.230}_{-0.230}$	&	$34.80^{+0.250}_{-0.250}$	&	650.182 	&	0.46 	&	0.77 	&	0.550 	&	0.870 	&	$	13.40 	$	&	$	-32.00 	$	&	400.20 	&	648.90 	&	495.30 	&	I	&	$<0.60$	\\
FRB20190601C	&	1	&	$88.52^{+0.180}_{-0.180}$	&	$28.47^{+0.057}_{-0.057}$	&	424.066 	&	1.32 	&	5.80 	&	0.510 	&	0.119 	&	$	36.10 	$	&	$	-79.50 	$	&	423.60 	&	595.40 	&	502.20 	&	I	&	$<0.60$	\\
FRB20190621C	&	0	&	$206.57^{+0.210}_{-0.210}$	&	$5.23^{+0.290}_{-0.290}$	&	570.267 	&	1.98 	&	2.38 	&	0.443 	&	0.510 	&	$	39.10 	$	&	$	-101.00 	$	&	417.50 	&	564.40 	&	485.40 	&	I	&	$<0.60$	\\
FRB20190410B	&	0	&	$265.76^{+0.020}_{-0.020}$	&	$15.17^{+0.200}_{-0.200}$	&	642.170 	&	0.22 	&	0.45 	&	0.423 	&	0.211 	&	$	18.70 	$	&	$	-73.40 	$	&	400.20 	&	542.50 	&	454.40 	&	I	&	$<0.60$	\\
FRB20190515D	&	0	&	$67.13^{+0.210}_{-0.210}$	&	$-5.01^{+0.340}_{-0.340}$	&	426.061 	&	3.00 	&	8.80 	&	0.420 	&	1.490 	&	$	30.60 	$	&	$	-54.30 	$	&	431.60 	&	651.50 	&	530.20 	&	I	&	$<0.60$	\\
FRB20190417C	&	0	&	$45.68^{+0.030}_{-0.030}$	&	$71.26^{+0.046}_{-0.046}$	&	320.232 	&	7.90 	&	10.80 	&	0.413 	&	0.430 	&	$	24.78 	$	&	$	-32.41 	$	&	449.30 	&	765.80 	&	586.60 	&	I	&	$<0.60$	\\
FRB20190208C	&	0	&	$141.55^{+0.037}_{-0.037}$	&	$83.56^{+0.220}_{-0.220}$	&	238.392 	&	1.27 	&	1.74 	&	0.411 	&	0.450 	&	$	18.70 	$	&	$	-54.60 	$	&	400.20 	&	583.20 	&	474.90 	&	I	&	$<0.60$	\\
FRB20181126A	&	0	&	$262.05^{+0.180}_{-0.180}$	&	$81.17^{+0.190}_{-0.190}$	&	494.217 	&	3.50 	&	9.40 	&	0.407 	&	0.165 	&	$	17.93 	$	&	$	-90.70 	$	&	400.20 	&	518.10 	&	441.80 	&	I	&	$<0.60$	\\
FRB20181030E	&	0	&	$135.67^{+0.012}_{-0.012}$	&	$8.89^{+0.190}_{-0.190}$	&	159.690 	&	2.00 	&	6.30 	&	0.400 	&	0.973 	&	$	22.30 	$	&	$	-69.00 	$	&	400.20 	&	564.80 	&	470.50 	&	I	&	$<0.60$	\\
FRB20190308C	&	0	&	$188.36^{+0.026}_{-0.026}$	&	$44.39^{+0.170}_{-0.170}$	&	500.519 	&	0.47 	&	4.80 	&	0.400 	&	2.290 	&	$	15.20 	$	&	$	-61.00 	$	&	400.20 	&	550.70 	&	453.40 	&	I	&	$<0.60$	\\
FRB20190426A	&	0	&	$115.04^{+0.200}_{-0.200}$	&	$59.12^{+0.200}_{-0.200}$	&	340.662 	&	1.59 	&	2.01 	&	0.398 	&	0.430 	&	$	27.00 	$	&	$	-71.70 	$	&	403.70 	&	577.80 	&	483.00 	&	I	&	$<0.60$	\\
FRB20190131E	&	0	&	$195.65^{+0.044}_{-0.044}$	&	$80.92^{+0.270}_{-0.270}$	&	279.801 	&	3.00 	&	5.10 	&	0.230 	&	0.164 	&	$	22.00 	$	&	$	-63.10 	$	&	400.20 	&	576.80 	&	476.50 	&	I	&	$<0.60$	\\
FRB20190308B	&	0	&	$38.59^{+0.040}_{-0.040}$	&	$83.62^{+0.300}_{-0.300}$	&	180.180 	&	1.11 	&	1.39 	&	0.186 	&	0.134 	&	$	18.60 	$	&	$	-52.90 	$	&	400.20 	&	587.90 	&	477.20 	&	I	&	$<0.60$	\\
FRB20180923A	&	0	&	$327.61^{+0.038}_{-0.038}$	&	$71.92^{+0.200}_{-0.200}$	&	219.440 	&	0.76 	&	1.20 	&	0.150 	&	0.211 	&	$	18.20 	$	&	$	-57.40 	$	&	400.20 	&	572.90 	&	468.90 	&	I	&	$<0.60$	\\
FRB20190118A	&	0	&	$253.31^{+0.029}_{-0.029}$	&	$11.55^{+0.084}_{-0.084}$	&	225.108 	&	9.30 	&	18.00 	&	0.140 	&	0.282 	&	$	19.42 	$	&	$	-56.71 	$	&	400.20 	&	580.90 	&	474.90 	&	I	&	$<0.60$	\\
FRB20180729A	&	0	&	$199.40^{+0.120}_{-0.120}$	&	$55.58^{+0.084}_{-0.084}$	&	109.594 	&	11.70 	&	17.00 	&	0.100 	&	0.157 	&	$	16.46 	$	&	$	-30.21 	$	&	400.20 	&	692.70 	&	525.60 	&	I	&	$<0.60$	\\
FRB20180923D	&	0	&	$169.08^{+0.020}_{-0.020}$	&	$48.75^{+0.070}_{-0.070}$	&	329.400 	&	2.40 	&	2.20 	&	0.100 	&	0.114 	&	$	20.90 	$	&	$	-92.70 	$	&	400.20 	&	524.40 	&	448.00 	&	I	&	$<0.60$	\\
FRB20181128C	&	1	&	$268.77^{+0.023}_{-0.023}$	&	$49.71^{+0.200}_{-0.200}$	&	618.350 	&	0.39 	&	3.40 	&	0.100 	&	2.320 	&	$	23.30 	$	&	$	-60.00 	$	&	400.20 	&	590.80 	&	485.80 	&	I	&	$<0.60$	\\
FRB20190110A	&	0	&	$64.95^{+0.033}_{-0.033}$	&	$47.44^{+0.086}_{-0.086}$	&	472.753 	&	1.54 	&	3.80 	&	0.203 	&	0.381 	&	$	6.30 	$	&	$	-118.00 	$	&	400.20 	&	472.70 	&	411.00 	&	II	&	0.70 	\\
FRB20190617B	&	0	&	$56.43^{+0.020}_{-0.020}$	&	$1.16^{+0.270}_{-0.270}$	&	273.510 	&	0.99 	&	9.20 	&	7.580 	&	9.500 	&	$	10.60 	$	&	$	-38.40 	$	&	400.20 	&	586.70 	&	459.30 	&	II	&	$<0.60$	\\
FRB20190601B	&	0	&	$17.88^{+0.170}_{-0.170}$	&	$23.82^{+0.036}_{-0.036}$	&	787.795 	&	1.00 	&	13.00 	&	4.040 	&	5.670 	&	$	9.70 	$	&	$	-68.60 	$	&	400.20 	&	515.90 	&	429.50 	&	II	&	$<0.60$	\\
FRB20181222D	&	0	&	$188.20^{+0.230}_{-0.230}$	&	$56.16^{+0.084}_{-0.084}$	&	1417.110 	&	0.22 	&	1.23 	&	3.750 	&	1.950 	&	$	8.40 	$	&	$	-41.10 	$	&	400.20 	&	561.30 	&	443.00 	&	II	&	$<0.60$	\\
FRB20190430A	&	0	&	$77.70^{+0.240}_{-0.240}$	&	$87.01^{+0.320}_{-0.320}$	&	339.250 	&	0.75 	&	7.70 	&	3.380 	&	3.230 	&	$	4.70 	$	&	$	-29.10 	$	&	400.20 	&	574.60 	&	433.80 	&	II	&	$<0.60$	\\
FRB20181214F	&	0	&	$252.62^{+0.047}_{-0.047}$	&	$32.44^{+0.220}_{-0.220}$	&	2105.760 	&	0.31 	&	2.21 	&	2.300 	&	1.520 	&	$	8.00 	$	&	$	-65.00 	$	&	400.20 	&	514.10 	&	425.70 	&	II	&	$<0.60$	\\
FRB20190224A	&	0	&	$60.53^{+0.250}_{-0.250}$	&	$83.39^{+0.290}_{-0.290}$	&	818.400 	&	0.63 	&	8.50 	&	2.040 	&	5.770 	&	$	2.60 	$	&	$	-60.00 	$	&	400.20 	&	497.50 	&	408.90 	&	II	&	$<0.60$	\\
FRB20190419A	&	0	&	$104.98^{+0.210}_{-0.210}$	&	$64.88^{+0.071}_{-0.071}$	&	439.972 	&	0.41 	&	0.77 	&	1.850 	&	2.400 	&	$	1.00 	$	&	$	-29.00 	$	&	400.20 	&	538.60 	&	407.10 	&	II	&	$<0.60$	\\
FRB20181125A	&	2	&	$147.94^{+0.180}_{-0.180}$	&	$33.93^{+0.041}_{-0.041}$	&	272.190 	&	0.39 	&	3.20 	&	1.580 	&	1.500 	&	$	7.30 	$	&	$	-58.00 	$	&	400.20 	&	520.90 	&	426.50 	&	II	&	$<0.60$	\\
FRB20181125A	&	1	&	$147.94^{+0.180}_{-0.180}$	&	$33.93^{+0.041}_{-0.041}$	&	272.190 	&	0.39 	&	3.20 	&	1.440 	&	1.500 	&	$	7.30 	$	&	$	-41.80 	$	&	400.20 	&	552.00 	&	436.60 	&	II	&	$<0.60$	\\
FRB20190114A	&	0	&	$8.95^{+0.230}_{-0.230}$	&	$19.17^{+0.260}_{-0.260}$	&	887.392 	&	0.55 	&	2.30 	&	1.340 	&	0.380 	&	$	11.10 	$	&	$	-89.00 	$	&	400.20 	&	500.00 	&	425.80 	&	II	&	$<0.60$	\\
FRB20190226C	&	0	&	$17.46^{+0.210}_{-0.210}$	&	$26.76^{+0.056}_{-0.056}$	&	827.770 	&	0.39 	&	1.41 	&	1.310 	&	1.500 	&	$	6.60 	$	&	$	-42.00 	$	&	400.20 	&	548.30 	&	433.30 	&	II	&	$<0.60$	\\
FRB20190614A	&	0	&	$179.79^{+0.091}_{-0.091}$	&	$88.33^{+0.240}_{-0.240}$	&	1064.039 	&	0.83 	&	2.21 	&	1.300 	&	0.770 	&	$	1.60 	$	&	$	-29.00 	$	&	400.20 	&	544.00 	&	411.10 	&	II	&	$<0.60$	\\
FRB20181117B	&	1	&	$81.09^{+0.190}_{-0.190}$	&	$79.99^{+0.099}_{-0.099}$	&	538.200 	&	3.60 	&	11.00 	&	1.280 	&	1.300 	&	$	11.40 	$	&	$	-31.20 	$	&	400.20 	&	630.40 	&	480.40 	&	II	&	$<0.60$	\\
FRB20181125A	&	0	&	$147.94^{+0.180}_{-0.180}$	&	$33.93^{+0.041}_{-0.041}$	&	272.190 	&	0.39 	&	3.20 	&	1.280 	&	1.500 	&	$	9.00 	$	&	$	-54.00 	$	&	400.20 	&	533.70 	&	434.50 	&	II	&	$<0.60$	\\
FRB20190222C	&	0	&	$239.18^{+0.040}_{-0.040}$	&	$40.03^{+0.190}_{-0.190}$	&	524.007 	&	0.44 	&	0.83 	&	0.676 	&	0.740 	&	$	9.30 	$	&	$	-53.90 	$	&	400.20 	&	536.40 	&	436.30 	&	II	&	$<0.60$	\\
FRB20190520A	&	0	&	$273.52^{+0.026}_{-0.026}$	&	$26.32^{+0.094}_{-0.094}$	&	432.506 	&	1.08 	&	2.40 	&	0.649 	&	0.710 	&	$	11.40 	$	&	$	-54.30 	$	&	400.20 	&	546.00 	&	444.40 	&	II	&	$<0.60$	\\
FRB20190618A	&	0	&	$321.25^{+0.170}_{-0.170}$	&	$25.44^{+0.075}_{-0.075}$	&	228.947 	&	2.40 	&	4.30 	&	0.548 	&	0.580 	&	$	3.34 	$	&	$	-35.80 	$	&	400.20 	&	540.40 	&	419.30 	&	II	&	$<0.60$	\\
FRB20190308B	&	1	&	$38.59^{+0.040}_{-0.040}$	&	$83.62^{+0.300}_{-0.300}$	&	180.180 	&	1.11 	&	1.39 	&	0.520 	&	0.134 	&	$	9.50 	$	&	$	-37.00 	$	&	400.20 	&	585.30 	&	455.50 	&	II	&	$<0.60$	\\
FRB20190605C	&	0	&	$168.32^{+0.048}_{-0.048}$	&	$-5.19^{+0.093}_{-0.093}$	&	187.642 	&	4.60 	&	4.40 	&	0.495 	&	0.520 	&	$	9.90 	$	&	$	-40.50 	$	&	400.20 	&	573.90 	&	452.20 	&	II	&	$<0.60$	\\
FRB20181130A	&	0	&	$355.19^{+0.031}_{-0.031}$	&	$46.49^{+0.170}_{-0.170}$	&	220.090 	&	0.97 	&	1.27 	&	0.480 	&	0.510 	&	$	2.70 	$	&	$	-45.70 	$	&	400.20 	&	515.90 	&	412.10 	&	II	&	$<0.60$	\\
FRB20190423A	&	0	&	$179.68^{+0.180}_{-0.180}$	&	$55.25^{+0.160}_{-0.160}$	&	242.646 	&	10.80 	&	55.40 	&	0.414 	&	0.539 	&	$	11.19 	$	&	$	-30.52 	$	&	400.20 	&	632.60 	&	480.70 	&	II	&	$<0.60$	\\
FRB20190304A	&	0	&	$124.51^{+0.036}_{-0.036}$	&	$74.61^{+0.210}_{-0.210}$	&	483.727 	&	0.71 	&	2.90 	&	0.408 	&	0.901 	&	$	6.30 	$	&	$	-67.30 	$	&	400.20 	&	504.70 	&	419.40 	&	II	&	$<0.60$	\\
FRB20181022E	&	0	&	$221.18^{+0.210}_{-0.210}$	&	$27.13^{+0.220}_{-0.220}$	&	285.986 	&	0.69 	&	2.08 	&	0.400 	&	0.632 	&	$	8.70 	$	&	$	-42.30 	$	&	400.20 	&	560.30 	&	443.70 	&	II	&	$<0.60$	\\
FRB20190517C	&	0	&	$87.50^{+0.190}_{-0.190}$	&	$26.62^{+0.200}_{-0.200}$	&	335.575 	&	3.10 	&	8.70 	&	0.377 	&	0.149 	&	$	8.48 	$	&	$	-50.20 	$	&	400.20 	&	539.40 	&	435.50 	&	II	&	$<0.60$	\\
FRB20190301D	&	0	&	$278.72^{+0.220}_{-0.220}$	&	$74.68^{+0.093}_{-0.093}$	&	1160.692 	&	0.39 	&	1.50 	&	0.360 	&	0.535 	&	$	3.60 	$	&	$	-29.40 	$	&	400.20 	&	562.70 	&	425.40 	&	II	&	$<0.60$	\\
FRB20190109B	&	0	&	$253.47^{+0.230}_{-0.230}$	&	$1.25^{+0.330}_{-0.330}$	&	175.168 	&	1.20 	&	3.00 	&	0.340 	&	0.261 	&	$	2.50 	$	&	$	-65.00 	$	&	400.20 	&	492.40 	&	408.10 	&	II	&	$<0.60$	\\
FRB20180928A	&	0	&	$312.95^{+0.040}_{-0.040}$	&	$30.85^{+0.053}_{-0.053}$	&	252.768 	&	1.34 	&	2.50 	&	0.269 	&	0.148 	&	$	-2.93 	$	&	$	-39.70 	$	&	400.20 	&	492.10 	&	400.20 	&	II	&	$<0.60$	\\
FRB20190621D	&	0	&	$270.63^{+0.016}_{-0.016}$	&	$78.89^{+0.170}_{-0.170}$	&	647.514 	&	0.89 	&	4.30 	&	2.290 	&	2.600 	&	$	7.40 	$	&	$	-22.30 	$	&	400.20 	&	651.10 	&	472.10 	&	III	&	$<0.60$	\\
FRB20190204A	&	0	&	$161.33^{+0.230}_{-0.230}$	&	$61.53^{+0.240}_{-0.240}$	&	449.639 	&	0.24 	&	1.50 	&	1.810 	&	0.880 	&	$	2.40 	$	&	$	-28.00 	$	&	400.20 	&	557.90 	&	418.20 	&	III	&	$<0.60$	\\
FRB20190701D	&	0	&	$112.10^{+0.180}_{-0.180}$	&	$66.70^{+0.160}_{-0.160}$	&	933.363 	&	1.33 	&	8.60 	&	1.400 	&	1.530 	&	$	6.49 	$	&	$	-20.90 	$	&	400.20 	&	651.80 	&	467.60 	&	III	&	$<0.60$	\\
FRB20190221A	&	0	&	$132.60^{+0.050}_{-0.050}$	&	$9.90^{+0.260}_{-0.260}$	&	223.806 	&	1.23 	&	2.33 	&	0.970 	&	0.408 	&	$	5.10 	$	&	$	-23.90 	$	&	400.20 	&	606.50 	&	444.80 	&	III	&	$<0.60$	\\
FRB20181222E	&	1	&	$50.64^{+0.044}_{-0.044}$	&	$86.97^{+0.250}_{-0.250}$	&	327.980 	&	1.12 	&	5.50 	&	0.910 	&	0.790 	&	$	9.90 	$	&	$	-23.80 	$	&	400.20 	&	672.30 	&	492.80 	&	III	&	$<0.60$	\\
FRB20190416B	&	0	&	$172.19^{+0.180}_{-0.180}$	&	$35.95^{+0.034}_{-0.034}$	&	575.360 	&	0.69 	&	1.47 	&	0.790 	&	0.491 	&	$	-3.60 	$	&	$	-23.50 	$	&	400.20 	&	511.90 	&	400.20 	&	III	&	$<0.60$	\\
FRB20190627A	&	0	&	$195.89^{+0.240}_{-0.240}$	&	$0.75^{+0.370}_{-0.370}$	&	404.221 	&	1.98 	&	2.62 	&	0.663 	&	0.800 	&	$	9.30 	$	&	$	-27.30 	$	&	400.20 	&	634.60 	&	474.60 	&	III	&	$<0.60$	\\
FRB20181030C	&	0	&	$309.83^{+0.220}_{-0.220}$	&	$3.99^{+0.300}_{-0.300}$	&	668.760 	&	1.60 	&	5.50 	&	0.610 	&	0.740 	&	$	10.00 	$	&	$	-22.20 	$	&	400.20 	&	691.20 	&	500.90 	&	III	&	$<0.60$	\\
FRB20181209A	&	0	&	$98.16^{+0.210}_{-0.210}$	&	$68.69^{+0.210}_{-0.210}$	&	328.656 	&	2.50 	&	3.20 	&	0.597 	&	0.640 	&	$	0.42 	$	&	$	-25.50 	$	&	400.20 	&	544.70 	&	403.50 	&	III	&	$<0.60$	\\
FRB20190502C	&	0	&	$155.60^{+0.150}_{-0.150}$	&	$82.97^{+0.040}_{-0.040}$	&	396.835 	&	3.60 	&	8.30 	&	0.527 	&	0.320 	&	$	9.00 	$	&	$	-26.80 	$	&	400.20 	&	634.30 	&	473.20 	&	III	&	$<0.60$	\\
FRB20190111A	&	1	&	$217.00^{+0.150}_{-0.150}$	&	$26.78^{+0.120}_{-0.120}$	&	171.968 	&	3.60 	&	17.00 	&	0.438 	&	0.535 	&	$	7.91 	$	&	$	-23.30 	$	&	400.20 	&	649.70 	&	474.40 	&	III	&	$<0.60$	\\
FRB20181222E	&	0	&	$50.64^{+0.044}_{-0.044}$	&	$86.97^{+0.250}_{-0.250}$	&	327.980 	&	1.12 	&	5.50 	&	0.254 	&	0.790 	&	$	5.13 	$	&	$	-19.90 	$	&	400.20 	&	639.50 	&	455.20 	&	III	&	$<0.60$	\\
FRB20190612B	&	0	&	$222.21^{+0.220}_{-0.220}$	&	$4.31^{+0.120}_{-0.120}$	&	187.600 	&	2.41 	&	3.78 	&	0.186 	&	0.121 	&	$	10.60 	$	&	$	-25.90 	$	&	400.20 	&	662.10 	&	491.30 	&	III	&	$<0.60$	\\
FRB20181230A	&	0	&	$346.69^{+0.220}_{-0.220}$	&	$83.37^{+0.260}_{-0.260}$	&	769.611 	&	0.94 	&	18.00 	&	1.640 	&	41.000 	&	$	33.00 	$	&	$	-27.80 	$	&	543.50 	&	800.20 	&	724.80 	&	IV	&	0.60 	\\
FRB20190206B	&	0	&	$49.76^{+0.250}_{-0.250}$	&	$79.50^{+0.390}_{-0.390}$	&	352.520 	&	0.95 	&	9.60 	&	7.100 	&	9.000 	&	$	11.60 	$	&	$	-24.60 	$	&	400.20 	&	687.60 	&	506.40 	&	IV	&	$<0.60$	\\
FRB20180920A	&	0	&	$78.89^{+0.210}_{-0.210}$	&	$28.29^{+0.230}_{-0.230}$	&	555.660 	&	0.86 	&	8.50 	&	2.220 	&	9.100 	&	$	20.10 	$	&	$	-26.30 	$	&	435.90 	&	787.40 	&	585.90 	&	IV	&	$<0.60$	\\
FRB20190411C	&	0	&	$9.33^{+0.054}_{-0.054}$	&	$20.50^{+0.210}_{-0.210}$	&	233.660 	&	3.19 	&	9.30 	&	1.023 	&	1.100 	&	$	24.30 	$	&	$	-26.10 	$	&	473.10 	&	800.20 	&	636.50 	&	IV	&	$<0.60$	\\
FRB20190213D	&	0	&	$336.45^{+0.021}_{-0.021}$	&	$52.71^{+0.140}_{-0.140}$	&	1346.848 	&	1.00 	&	2.20 	&	0.626 	&	0.700 	&	$	26.20 	$	&	$	-25.30 	$	&	496.60 	&	800.20 	&	671.40 	&	IV	&	$<0.60$	\\
FRB20190210D	&	0	&	$307.80^{+0.082}_{-0.082}$	&	$55.46^{+0.180}_{-0.180}$	&	359.148 	&	1.37 	&	2.50 	&	0.580 	&	0.273 	&	$	20.70 	$	&	$	-24.40 	$	&	449.90 	&	800.20 	&	611.80 	&	IV	&	$<0.60$	\\
FRB20180904A	&	0	&	$286.58^{+0.170}_{-0.170}$	&	$81.22^{+0.120}_{-0.120}$	&	361.137 	&	3.80 	&	6.00 	&	0.528 	&	0.550 	&	$	12.12 	$	&	$	-23.18 	$	&	400.20 	&	712.30 	&	519.70 	&	IV	&	$<0.60$	\\
\enddata
\tablecomments{%
The `Sub\_num' column represents the sub-burst number assigned to each FRB in the CHIME/FRB sample. The `Quadrant' column denotes the $\gamma$-$r$ quadrant to which each FRB belongs (see Section \ref{ssec:probesti} for details).
}
\end{deluxetable*}

\end{document}